\documentclass[conference]{IEEEtran}

\usepackage{times}
\usepackage{amssymb}

\usepackage{amsthm}
\usepackage{amsmath}
\usepackage{epsfig}
\usepackage{subcaption}
\usepackage[ruled, vlined]{algorithm2e}
\usepackage[para]{footmisc}
\usepackage{array}
\usepackage{url}
\usepackage{comment}
\usepackage{multicol}
\usepackage{balance}

\let\emptyset\varnothing

\newcommand{\eat}[1]{}
\newcommand{\attrname}[1]{\textsf{\footnotesize #1}}
\newcommand{\attrval}[1]{\textsl{\footnotesize #1}}
\newcommand{\attrnametwo}[1]{\textsf{\scriptsize #1}}

\newcommand{\algname}[1]{\textsf{\footnotesize #1}}

\newcommand{\reminder}[1]{ (((\mbox{$\Longleftarrow \star$}{\textbf{#1}} )))}

\renewcommand{\qed}{\hfill \mbox{\raggedright \rule[0pt]{1.2ex}{1.2ex}}}
\newcommand{\qedproof}{\hfill $\square$}

{
\theoremstyle{definition}
\newtheorem{definition}{Definition}
\newtheorem{proposition}{Proposition}

\newtheorem{example}{Example}

\newtheorem{invariant}{Invariant}
}

\makeatletter
\newcommand\semiHuge{\@setfontsize\semiHuge{23.72}{27.38}}
\makeatother

\begin{document}\sloppy


\title{\semiHuge Incremental Discovery of Prominent Situational Facts}
\author{
{Afroza Sultana{\small $~^{1}$}, Naeemul Hassan{\small $~^{1}$}, Chengkai Li{\small $~^{1}$}, Jun Yang{\small $~^{2}$}, Cong Yu{\small $~^{3}$} }%
\vspace{1.6mm}\\
\fontsize{10}{10}\selectfont\itshape
$^{1}$University of Texas at Arlington, $^{2}$Duke University, $^{3}$Google Research
}

\maketitle
\thispagestyle{plain}
\pagestyle{plain}


\begin{abstract}
  We study the novel problem of finding new, \emph{prominent
    situational facts}, which are emerging statements about objects
  that stand out within certain contexts. Many such facts are
  newsworthy---e.g., an athlete's outstanding performance in a game,
  or a viral video's impressive popularity. Effective and efficient
  identification of these facts assists journalists in reporting, one
  of the main goals of computational journalism.  Technically, we
  consider an ever-growing table of objects with \emph{dimension} and
  \emph{measure} attributes.  A situational fact is a ``contextual''
  skyline tuple that stands out against historical tuples in a
  context, specified by a conjunctive constraint involving dimension
  attributes, when a set of measure attributes are compared. New
  tuples are constantly added to the table, reflecting events
  happening in the real world.  Our goal is to discover
  constraint-measure pairs that qualify a new tuple as a contextual
  skyline tuple, and discover them quickly before the event becomes
  yesterday's news.  A brute-force approach requires exhaustive
  comparison with every tuple, under every constraint, and in every
  measure subspace.  We design algorithms in response to these
  challenges using three corresponding ideas---tuple reduction,
  constraint pruning, and sharing computation across measure
  subspaces.  We also adopt a simple prominence measure to rank the
  discovered facts when they are numerous.  Experiments over two real
  datasets validate the effectiveness and efficiency of our
  techniques.
\end{abstract}


\section{Introduction}\label{sec:intro}

\emph{Computational journalism} emerged recently as a young
interdisciplinary field~\cite{compjour-cacm} that brings together
experts in journalism, social sciences and computer science, and
advances journalism by innovations in computational techniques.
Database and data mining researchers have also
started to push the frontiers of this
field~\cite{compjour-cidr,streak-kdd11,oneofthefew-kdd12}.  One of the
goals in computational journalism is \emph{newsworthy fact discovery}.
Reporters always try hard to bring out attention-seizing factual
statements backed by data, which may lead to news stories and
investigation.  While such statements take many different forms, we
consider a common form exemplified by the following excerpts from
real-world news media:
\begin{list}{$\bullet$}
{ \setlength{\leftmargin}{1em} \setlength{\itemsep}{-1pt} }
\item ``Paul George had 21 points, 11 rebounds and 5 assists to become
the first Pacers player with a 20/10/5 (points/rebounds/assists) game against the Bulls
since Detlef Schrempf in December 1992.''
({\small \url{http://espn.go.com/espn/elias?date=20130205}})
\item ``The social world's most viral photo ever generated 3.5 million likes,
170,000 comments and 460,000 shares by Wednesday afternoon.''
({\small \url{http://www.cnbc.com/id/49728455}})
\end{list}

What is common in the above two statements is a prominent fact with regard to a context and several measures.
In the first statement, the context includes the performance of Pacers players in games against the Bulls
since December 1992 and the measures are \attrname{points}, \attrname{rebounds}, \attrname{assists}.  By
these measures, no performance in the context is better than the mentioned performance of Paul George.  For
the second statement, the measures are \attrname{likes}, \attrname{comments}, \attrname{shares} and the
context includes all photos posted to Facebook.  The story is that no photo in the context attracted more
attention than the mentioned photo of President Barack Obama, by the three measures.  In general, facts can
be put in many contexts, such as photos posted in 2012, photos posted by political campaigns, and so on.

Similar facts can be stated on data from domains outside of sports and
social media, including stock data, weather data, and criminal
records.  For example: 1)~``Stock A becomes the first stock in history
with price over \$300 and market cap over \$400 billion.''
2)~``Today's measures of wind speed and humidity are $x$ and $y$,
respectively.  City B has never encountered such high wind speed and
humidity in March.''  3)~``There were 35 DUI arrests and 20 collisions
in city C yesterday, the first time in 2013.''  Some of these facts
are not only interesting to reporters but also useful to financial
analysts, scientists, and citizens.

In technical terms, a fact considered in this paper is a \emph{contextual skyline} object that stands out
against other objects in a context with regard to a set of measures.  Consider a table $R$ whose schema
includes a set of measure attributes $\mathcal{M}$ and a set of dimension attributes $\mathcal{D}$.  A
context is a subset of $R$, resulting from a conjunctive constraint defined on a subset of the dimension
attributes $D \subseteq \mathcal{D}$.  A measure subspace is defined by a subset of the measure attributes
$M \subseteq \mathcal{M}$.  A tuple $t$ is a contextual skyline tuple if no other tuple in the context
dominates $t$.  A tuple $t'$ dominates $t$ if $t'$ is better than or equal to $t$ on every attribute in $M$
and better than $t$ on at least one of the attributes.  Such is the standard notion of dominance relation
adopted in \emph{skyline analysis}~\cite{656550}.

We study how to find \emph{situational facts} pertinent to new tuples in an ever-growing database,
where the tuples capture real-world events.  We propose algorithms that, whenever a new tuple $t$ enters
an append-only table $R$, discover constraint-measure pairs that qualify $t$ as a
contextual skyline tuple. Each such pair constitutes a situational fact
pertinent to $t$'s arrival.

\begin{table*}[t]
      \begin{center}
          \scriptsize
          \vspace{-3mm}
            \begin{tabular}{|*{10}{c|}}\hline
            \attrname{tuple id} &\attrname{player}& \attrname{day} & \attrname{month} & \attrname{season} & \attrname{team} & \attrname{opp\_team} & \attrname{points} & \attrname{assists} & \attrname{rebounds}\\
            \hline
            \hline
            $t_1$ & Bogues & 11 & Feb. & 1991-92 & Hornets & Hawks & 4 & 12 & 5\\
            \hline
            $t_2$ & Seikaly & 13 & Feb. & 1991-92 & Heat & Hawks & 24 & 5 & 15\\
            \hline
            $t_3$ & Sherman & 7 & Dec. & 1993-94 & Celtics & Nets & 13 & 13 & 5\\
            \hline
            $t_4$ & Wesley & 4 & Feb. & 1994-95 & Celtics & Nets & 2 & 5 & 2\\
            \hline
            $t_5$ & Wesley & 5 & Feb. & 1994-95 & Celtics & Timberwolves & 3 & 5 & 3\\
            \hline
            $t_6$ & Strickland & 3 & Jan. & 1995-96 & Blazers & Celtics & 27 & 18 & 8\\
            \hline
            \hline
            $t_7$ & Wesley & 25 & Feb. & 1995-96 & Celtics & Nets & 12 & 13 & 5\\
            \hline
            \end{tabular}
      \end{center}
      \vspace{-1mm}\hspace{25mm}{* Attribute \attrname{opp\_team} is the short form of \attrname{opposition team}.}
      \vspace{-1mm}\caption{\small A Mini-world of Basketball Gamelogs}\vspace{-6mm}
      \label{tab:nba_example}
\end{table*}

\vspace{-1.5mm}
\begin{example}\label{ex:motivate}
Consider the mini-world of basketball gamelogs $R$ in Table~\ref{tab:nba_example},
where $\mathcal{D}$=\{\attrname{player}, \attrname{month}, \attrname{season},
\attrname{team}, \attrname{opp\_team}\} and $\mathcal{M}$=\{\attrname{points},
\attrname{assists}, \attrname{rebounds}\}.  The existing tuples are $t_1$ to
$t_6$ and the new tuple is $t_7$. If the context is the whole
table (i.e., no constraint) and the measure subspace $M$=$\mathcal{M}$,
$t_7$ is not a skyline tuple since it is dominated by $t_3$ and $t_6$.
However, with regard to context $\sigma_{\attrname{month}=\attrval{Feb.}}(R)$
(corresponding to constraint \attrname{month}=\attrval{Feb.}) and the same
measure subspace $M$, $t_7$ is in the skyline along with $t_2$.
In yet another context $\sigma_{\attrname{team}=\attrval{Celtics} \wedge}$
$_{\attrname{opp\_team}=\attrval{Nets}}(R)$ under measure subspace
$M$=\{\attrname{assists}, \attrname{rebounds}\}, $t_7$ is in the skyline
along with $t_3$. Tuple $t_7$ is also a contextual skyline tuple for other
constraint-measure pairs, which we do not further enumerate.\qed\vspace{-2mm}
\end{example}

Discovering situational facts is challenging as timely discovery of
such facts is expected.  In finding news leads centered around
situational facts, the value of a news piece diminishes rapidly after
the event takes place.  Consider NBA games again.  Sports media need
to identify and discuss sensational records quickly as they emerge;
any delay makes fans less interested in the records and risks
losing them to rival media.  Timely identification of situational
facts is also critical in areas beyond journalism.  To make informed
investment decisions, investors want to know facts related to stock
trading as soon as possible.  Facts discovered from weather
data can assist scientists in identifying extreme weather conditions
and help government and the public in coping with the weather.

Simple situational facts on a single measure and a complete
table, e.g., the all-time NBA scoring record, can be conveniently
detected by database triggers.  However, general and complex facts
involving multiple dimension and measure attributes are much harder to
discover.  Exhaustively using triggers leads to an exponential
explosion of constraint-measure pairs to check for each new
tuple.  In reality, news media relies on instincts and experiences of
domain experts on this endeavor.  The experts, impressed by an event
such as the outstanding performance of a player in a game, hypothesize
a fact and manually craft a database query to check it.  This is
how Elias Sports Bureau tackles the task and provides sports records
(such as the aforementioned one by Paul George) to many
sports media~\cite{hirdt}.  With ever-growing data and limited human
resources, such manual checking is time-consuming and error-prune.
Its low efficiency not only leads to delayed and
missing facts, but also ties up precious human expertise that could be
otherwise devoted to more important journalistic activities.

The technical focus of this paper is thus on efficient automatic
approach to discovering situational facts, i.e., finding
constraint-measure pairs that qualify a new tuple $t$ as a contextual
skyline tuple.  A straightforward brute-force approach would compare
$t$ with every historical tuple to determine if $t$ is dominated,
repeatedly for every conjunctive constraint satisfied by $t$
under every possible measure subspace.  The obvious low-efficiency of this approach has three
culprits---exhaustive comparison with \emph{every tuple},
under \emph{every constraint}, and over \emph{every measure subspace}.
We thus design algorithms to counter these issues by three
corresponding ideas, as follows:

\textbf{1) Tuple reduction}\hspace{2mm} Instead of comparing $t$ with
every previous tuple, it is sufficient to only compare with current
skyline tuples.  This is based on the simple property that, if any
tuple dominates $t$, then there must exist a skyline tuple that also
dominates $t$.  For example, in Table~\ref{tab:nba_example}, under
constraint \attrname{month}=\attrval{Feb.} and the full measure space
$\mathcal{M}$, the corresponding context contains $t_1$, $t_2$, $t_4$
and $t_5$, and the contextual skyline has two tuples---$t_1$ and
$t_2$.  When the new tuple $t_7$ comes, with regard to the same
constraint-measure pair, it suffices to compare $t_7$ with $t_1$ and
$t_2$, not the remaining tuples.

\textbf{2) Constraint pruning}\hspace{1mm} If $t$ is dominated by
$t'$ in a particular measure subspace $M$, then $t$ does not
belong to the contextual skyline of constraint-measure pair $(C,M)$
for any $C$ satisfied by both $t$ and $t'$.  For example,
since $t_7$ is dominated by $t_3$ in the full measure space $\mathcal{M}$,
it is not in the contextual skylines for
$(\attrname{team}$=$\attrval{Celtics} \wedge
\attrname{opp\_team}$=$\attrval{Nets},\mathcal{M})$,
$(\attrname{team}$=$\attrval{Celtics},\mathcal{M})$,
$(\attrname{opp\_team}$=$\attrval{Nets},\mathcal{M})$ and $($no
constraint$, \mathcal{M})$.  Furthermore, since $t_7$ is dominated by
$t_6$ in $\mathcal{M}$, it does not belong to the contextual skylines
for $(\attrname{season}$=$\attrval{1995-96},\mathcal{M})$ and
$($no constraint$, \mathcal{M})$.  Based on this, we
examine the constraints satisfied by $t$ in a certain order, such that
comparisons of $t$ with skyline tuples associated with already examined
constraints are used to prune remaining constraints from
consideration.

\textbf{3) Sharing computation across measure subspaces}\hspace{2mm}
Since repeatedly visiting the constraints satisfied by $t$ for every
measure subspace is wasteful, we pursue sharing computation across
different subspaces.  The challenge in such sharing lies in the
anti-monotonicity of dominance relation---a skyline tuple in space $M$
may or may not be in the skyline of a superspace or subspace $M'$~\cite{PeiYLJELWTYZ06}.
Nonetheless, we can
first consider the full space $\mathcal{M}$ and prune various
constraints from consideration for smaller subspaces.  For instance,
after comparing $t_7$ with $t_2$ in $\mathcal{M}$, the
algorithms realize that $t_7$ has smaller values on
\attrname{points} and \attrname{rebounds}.  It is dominated by $t_2$
in three subspaces---\{\attrname{points}, \attrname{rebounds}\},
\{\attrname{points}\} and \{\attrname{rebounds}\}.  When considering
these subspaces, we can skip two contexts---corresponding to
constraint \attrname{month}=\attrval{Feb.} and empty constraint,
respectively---as $t_2$ and $t_7$ are in both contexts.

It is crucial to report truly \emph{prominent} situational facts.  A
newly arrived tuple $t$ may be in the contextual skylines for many
constraint-measure pairs.  Reporting all of them will overwhelm users
and make important facts harder to spot.  We measure the
\emph{prominence} of a constraint-measure pair by the cardinality
ratio of all tuples to skyline tuples in the corresponding context.
The intuition is that, if $t$ is one of the very few skyline tuples in
a context containing many tuples under a measure subspace, then the
corresponding constraint-measure pair brings out a prominent fact.  We
thus rank all situational facts pertinent to $t$ in descending order
of prominence.  Reporters and experts can choose to investigate
top-\emph{k} facts or the facts with prominence values above a
threshold.

The contributions of this paper are summarized as follows:\vspace{-0.5mm}
\begin{list}{$\bullet$}
{ \setlength{\leftmargin}{0.5em}}
\item We study the novel problem of finding situational facts and formalize it as discovering constraint-measure pairs that qualify a tuple as a contextual skyline tuple.

\item We devise efficient algorithms based on three main ideas---tuple reduction, constraint pruning and sharing computation across measure subspaces.

\item We use a simple prominence measure for ranking situational facts and discovering prominent situational facts.

\item We conduct extensive experiments on two real datasets (NBA dataset and weather dataset) to investigate their prominent situational facts and to study the efficiency of various proposed algorithms and their tradeoffs.
\end{list}


\section{Related Work}\label{sec:related}
\vspace{-1.5mm}
Pioneers in data journalism have considerable success in using computer programs to write stories about sports games and stock earnings (e.g., StatSheet {\small \url{http://statsheet.com/}} and Narrative Science {\small \url{http://www.narrativescience.com/}}).  The stories follow writing patterns to narrate box scores and play-by-play data and a company's earnings data.  They focus on capturing what happened in the game or what the earnings numbers indicate.  They do not find situational facts pertinent to a game or an earnings report in the context of historical data.

Skyline query is extensively investigated in recent years, since B\"{o}rzs\"{o}nyi et al.~\cite{656550} brought the concept to the database field.  In~\cite{656550} and the studies afterwards, it is assumed both the context of tuples in comparison and the measure space are given as query conditions.  A high-level perspective on what distincts our work is---while prior studies \emph{find answers} (i.e., skyline points) for a given query (i.e., a context, a measure space, or their combination), we study the reverse problem of \emph{finding queries} (i.e., constraint-measure pairs that qualify a tuple as a contextual skyline tuple, among all possible pairs) for a particular answer (i.e., a new tuple).  

From a technical perspective, Table~\ref{tab:related} summarizes the differences among the more relevant previous studies and this paper, along three aspects---whether they consider all possible contexts defined on dimension attributes, all measure subspaces, and incremental computation on dynamic data.
With regard to context, Zhang et al.~\cite{Zhang10skylineconstraint} integrate the evaluation of a constraint with finding skyline tuples in the corresponding context in a given measure space.  With regard to measure, Pei et al.~\cite{PeiYLJELWTYZ06} compute on static data the \emph{skycube}---skyline points in all measure subspaces.  Xia et al.~\cite{xia2006refreshing} studied how to update a compressed skycube (CSC) when data change.  The CSC stores a tuple $t$ in its \emph{minimum subspaces}---the measure subspaces in which $t$ is a skyline tuple and of which the subspaces do not contain $t$ in the skyline.  They proposed an algorithm to update CSC when new tuples come and also an algorithm to use CSC to find all skyline tuples for a given measure subspace.

We can adapt \cite{xia2006refreshing} to find situational facts.  While Sec.~\ref{sec:exp} provides experimental comparisons with the adaptation, here we analyze its shortcomings.  Since \cite{xia2006refreshing} does not consider different contexts, the adaptation entails maintaining a separate CSC for every possible context.  Upon the arrival of a new tuple $t$, for every context, the adaptation will update the corresponding CSC.  Since a CSC only stores $t$ in its minimum subspaces, the adaptation needs to run their query algorithm to find the skyline tuples for all measure subspaces, in order to determine if $t$ is one of the skyline tuples.  This is clearly an overkill, caused by that CSC is designed for finding all skyline tuples.  Furthermore, while our algorithms can share computation across measure subspaces, there does not appear to be an effective strategy to share the computation of CSC algorithms across different contexts.

\emph{Promotion analysis by ranking}~\cite{Wu:2009:PAM:1687627.1687641} finds the contexts 
in which an object is ranked high.  
It ranks objects by a single score attribute, while we define object dominance relation on multiple measure attributes. 
It considers one-shot computation on static data, while we focus on incremental discovery on dynamic data.  Due to these distinctions, the algorithmic approaches in the two works are also fundamentally different.

Wu et al.~\cite{oneofthefew-kdd12} studied the \emph{one-of-the-$\tau$ object} problem, which entails finding the largest $k$ value and the corresponding $k$-\emph{skyband objects} (objects dominated by less than $k$ other objects) such that there are no more than $\tau$ $k$-skyband objects.  They consider all measure subspaces but not different contexts formed by constraints.  Similar to~\cite{Wu:2009:PAM:1687627.1687641}, it focuses on static data.

Alvanaki et al.~\cite{alvanaki2013interesting} worked on detecting interesting events through monitoring changes in ranking, by using materialized view maintenance techniques.  The work focuses on top-$k$ queries on single ranking attribute rather than skyline queries defined on multiple measure attributes.  Their ranking contexts have at most three constraints.  The work is similar to~\cite{avinash} which studied how to predict significant events based on historical data and correspondingly perform lazy maintenance of ranking views on a database.

\begin{table}[t]
\centering
\scriptsize
\begin{tabular}{|l|*{3}{c|}}\hline
 & all possible contexts & measure subspaces & incremental \\
\hline
~\cite{Zhang10skylineconstraint} & no & no & no \\
\hline
~\cite{PeiYLJELWTYZ06} & no & yes & no \\
\hline
~\cite{xia2006refreshing} & no & yes & yes \\
\hline
~\cite{Wu:2009:PAM:1687627.1687641} & yes & no & no \\
\hline
~\cite{oneofthefew-kdd12} & no & yes & no \\
\hline
~\cite{alvanaki2013interesting} & no & no & yes \\
\hline
this work& yes & yes & yes \\
\hline
\end{tabular}
\vspace{-2mm}\caption{\small Comparing Related Work on Three Modeling Aspects}\vspace{-3mm}
\label{tab:related}
\end{table}


\begin{table}[t]
\scriptsize
\centering
\begin{tabular}{|@{\hspace{0.3em}}c@{\hspace{0.3em}}|@{\hspace{0.3em}}l@{\hspace{0.3em}}|}
\hline
$R(\mathcal{D};\mathcal{M})$ & relation $R$, dimension space $\mathcal{D}$, measure space $\mathcal{M}$\\
\hline
$D \subseteq \mathcal{D}$ & dimension subspace\\
\hline
$M \subseteq \mathcal{M}$ & measure subspace\\
\hline
$C$ & constraint\\
\hline
($\mathcal{C}_\mathcal{D}$, $\trianglelefteq$) & poset of all constraints on subsumption relation $\trianglelefteq$\\
\hline
$C_1 \vartriangleleft (\trianglelefteq) C_2$ & $C_1$ is subsumed by (subsumed by or equal to) $C_2$\\
\hline
$t_1 \prec (\preceq) t_2$ & $t_1$ is dominated by (dominated by or equal to) $t_2$\\
\hline
$\sigma_{C}(R)$ & tuples in $R$ satisfying constraint $C$\\
\hline
$\lambda_{M}(R)$ & skyline tuples in $R$ on measure subspace $M$\\
\hline
$\lambda_{M}(\sigma_C$ $(R))$ & contextual skyline of $R$ with respect to $C$ and $M$\\
\hline
$\mu_{C,M}$ & tuples stored with respect to $C$ and $M$\\
\hline
$S^t$ & contextual skylines for $t$\\
\hline
$\mathcal{C}_{\mathcal{D}}^t$ or $\mathcal{C}^t$ & tuple-satisfied constraints of $t$\\
\hline
$\top$ & the top element of lattice ($\mathcal{C}_{\mathcal{D}}^t$, $\trianglelefteq$) and poset ($\mathcal{C}_{\mathcal{D}}$, $\trianglelefteq$)\\
\hline
$\bot(\mathcal{C}_{\mathcal{D}}^t)$ & the bottom element of lattice ($\mathcal{C}_{\mathcal{D}}^t$, $\trianglelefteq$) \\
\hline
$\mathcal{A}_C$,$\mathcal{D}_C$,$\mathcal{P}_C$,$\mathcal{CH}_C$ & $C$'s ancestors, descendants, parents, children in $\mathcal{C}_{\mathcal{D}}$\\
\hline
$\mathcal{A}^t_C$,$\mathcal{D}^t_C$,$\mathcal{P}^t_C$,$\mathcal{CH}^t_C$ & $C$'s ancestors, descendants, parents, children in $\mathcal{C}^t_{\mathcal{D}}$\\
\hline
$\mathcal{C}^{t_1,t_2}$ & the intersection of $\mathcal{C}^{t_1}$ and $\mathcal{C}^{t_2}$\\
\hline
$\mathcal{SC}^t_M$ & the skyline constraints of $t$ in $M$\\
\hline
$\mathcal{MSC}^t_M$ & the maximal skyline constraints of $t$ in $M$\\
\hline
\end{tabular}
\vspace{-1mm}\caption{Notations}\label{tab:notation}\vspace{-2mm}
\end{table}

\begin{table}[t]
\centering
\scriptsize
\begin{tabular}{|l|*{5}{c|}}\hline
$id$ & $d_1$ & $d_2$ & $d_3$ & $m_1$ & $m_2$\\
\hline
\hline
$t_1$ & $a_1$ & $b_2$ & $c_2$ & 10 & 15\\
\hline
$t_2$ & $a_1$ & $b_1$ & $c_1$ & 15 & 10\\
\hline
$t_3$ & $a_2$ & $b_1$ & $c_2$ & 17 & 17\\
\hline
$t_4$ & $a_2$ & $b_1$ & $c_1$ & 20 & 20\\
\hline
\hline
$t_5$ & $a_1$ & $b_1$ & $c_1$ & 11 & 15\\
\hline
\end{tabular}
\vspace{-1mm}\caption{\small Running Example}\vspace{-1mm}
\label{tab:running_example}
\end{table}

\section{Problem Statement}\label{sec:model}
\vspace{-1.5mm}
This section provides a formal description of our data model and problem statement. Table~\ref{tab:notation} lists the major notations. 
Consider a relational schema $R(\mathcal{D};\mathcal{M})$, where the
\emph{dimension space} is a set of \emph{dimension attributes}
$\mathcal{D}$=$\{d_1,\ldots$ $,d_n\}$ on which \emph{constraints} are
specified, and the \emph{measure space} is a set of \emph{measure attributes}
$\mathcal{M}$=$\{m_1,\ldots,m_s\}$ on which dominance relation for skyline
operation is defined.  Any set of dimension attributes
$D \subseteq \mathcal{D}$ defines a \emph {dimension subspace} and any
set of measure attributes $M \subseteq \mathcal{M}$ defines a
\emph{measure subspace}.  In Table~\ref{tab:running_example},
$R(\mathcal{D};\mathcal{M})$ = $\{t_1, t_2, t_3, t_4, t_5\}$,
$\mathcal{D}$ = $\{d_1, d_2, d_3\}$, $\mathcal{M}$=$\{m_1, m_2\}$.
We will use this table as a running example.\vspace{-1.5mm}

\begin{definition}[Constraint]
A \emph{constraint} $C$ on dimension space $\mathcal{D}$ is a conjunctive expression of the form $d_1$=$v_1$$\wedge$$d_2$$=$$v_2$$\wedge$ $\ldots \wedge$ $d_n$$=$$v_n$ (also written as $\langle v_1, v_2, \ldots, v_n \rangle$ for simplicity), where $v_i$$\in$$dom(d_i)$$\cup$ $\{*\}$ and $dom(d_i)$ is the value domain of dimension attribute $d_i$.  We use $C.d_i$ to denote the value $v_i$ assigned to $d_i$ in $C$.  If $C.d_i$$=$$*$, we say $d_i$ is \emph{unbound}, i.e., no condition is specified on $d_i$. We denote the number of bound attributes in $C$ as $bound(C)$.

The set of all possible constraints over dimension space $\mathcal{D}$
is denoted $\mathcal{C}_{\mathcal{D}}$.  Clearly,
$|\mathcal{C}_{\mathcal{D}}| = \prod_{i}(|dom(d_i)|+1)$.

Given a constraint $C \in \mathcal{C}_{\mathcal{D}}$, $\sigma_C(R)$ is the relational algebra expression that chooses all tuples in $R$ that satisfy $C$.\qed\vspace{-2mm}
\end{definition}

\begin{example}
For Table~\ref{tab:running_example}, an example constraint is $C = \langle a_1, *, c_1 \rangle$ in which $d_2$ is unbound.  $\sigma_C(R) = \{t_2, t_5\}$.\qed\vspace{-2mm}
\end{example}

\begin{figure*}[htb]
\begin{minipage}{0.3\textwidth}
\vspace{-3mm}
\epsfig{file=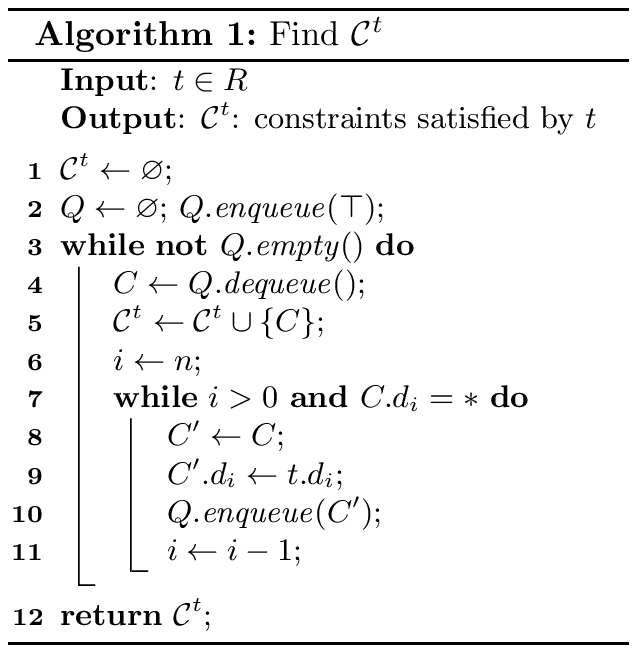, height=49mm,clip=}
\end{minipage}
\hspace{-2mm}
\begin{minipage}{0.35\textwidth}
\vspace{-3mm}
\epsfig{file=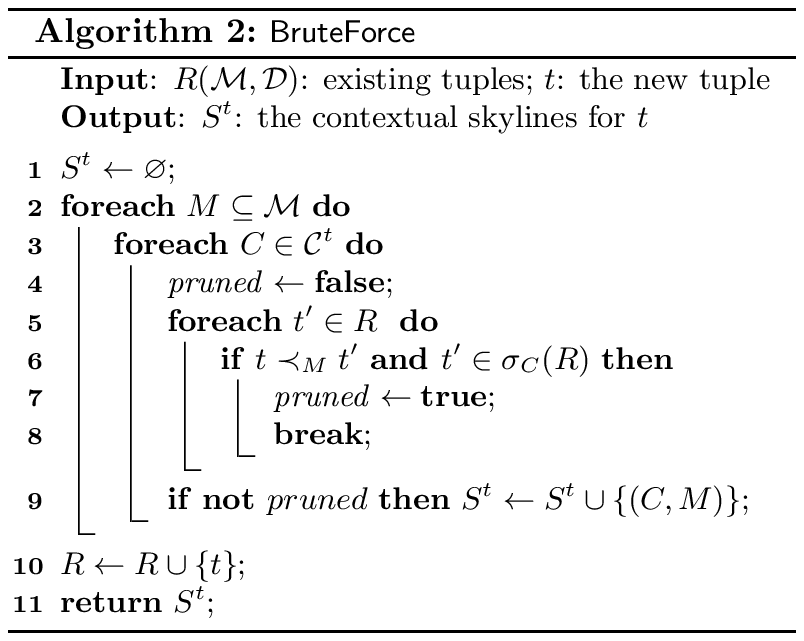, height=49mm,clip=}
\end{minipage}
\hspace{1mm}
\begin{minipage}{0.34\textwidth}
\vspace{-11mm}
\epsfig{file=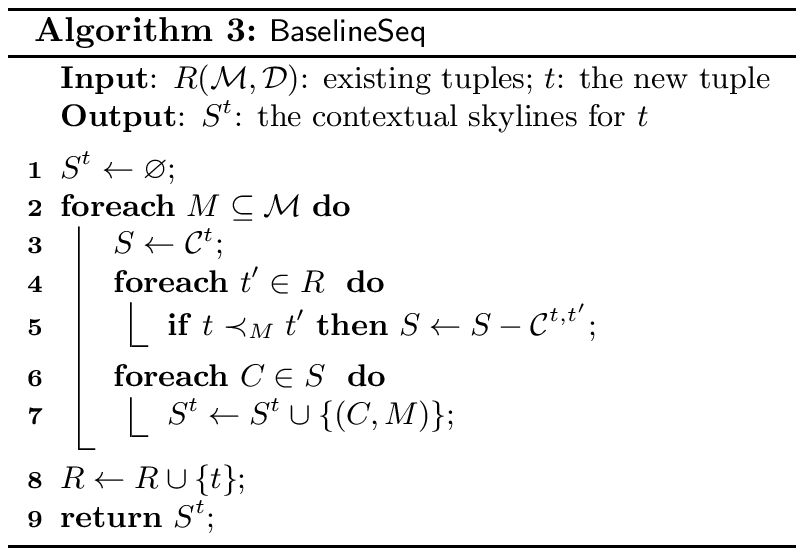, height=41mm,clip=}
\end{minipage}
\end{figure*}

\newcommand{\algfindconstraints}{1}
\newcommand{\algbruteforce}{2}
\newcommand{\algbaseline}{3}
\setcounter{algocf}{3}

\newcommand{\linewhileloop}{7}
\newcommand{\lineinitS}{3}
\newcommand{\linesubtract}{5}

\begin{definition}[Skyline]
\label{def:skyline}
Given a measure subspace $M$ and two tuples $t,t'\in R$, $t$ \emph{dominates} $t'$ with respect to $M$, denoted by $t$ $\succ_{M}$ $t'$ or $t'$ $\prec_{M}$ $t$, if $t$ is equal to or better than $t'$ on all attributes in $M$ and $t$ is better than $t'$ on at least one attribute in $M$.  A tuple $t$ is a \emph{skyline tuple} in subspace $M$ if it is not dominated by any other tuple in $R$. The set of all skyline tuples in $R$ with respect to $M$ is denoted by $\lambda_{M}(R)$, i.e., $\lambda_{M}(R)$=$\{t\in R|\nexists t'\in R\ \text{s.t.}\ t' \succ_M t\}$.\qed\vspace{-1.5mm}
\end{definition}

We use the general term ``better than'' in Def.~\ref{def:skyline}, which can mean either ``larger than'' or ``smaller than'' for numeric attributes and either ``ordered before'' or ``ordered after'' for ordinal attributes, depending on applications.  Further, the preferred ordering of values on different attributes are allowed to be different.  For example, in a basketball game, 10 points is better than 5 points, while 3 fouls is worse than 1 foul.  Without loss of generality, we assume measure attributes are numeric and a larger value is better than a smaller value.\vspace{-2mm} 

\begin{definition}[Contextual Skyline]
  Given a relation $R(\mathcal{D};\mathcal{M})$, the \emph{contextual
    skyline} under constraint $C$$\in$$\mathcal{C}_{\mathcal{D}}$ over
  measure subspace $M$$\subseteq$$\mathcal{M}$, denoted
  $\lambda_{M}(\sigma_C(R))$, is the skyline of $\sigma_C(R)$ in
  $M$.\qed\vspace{-2mm}
\end{definition}

\begin{example}
For Table~\ref{tab:running_example}, if $M = \mathcal{M}$,  $\lambda_{M}(R)$ = $\{t_4\}$. In fact, $t_4$ dominates all other tuples in space $M$.  If the constraint is $C = \langle a_1, b_1, c_1 \rangle$, $\sigma_C(R) = \{t_2, t_5\}$, $\lambda_{M}(\sigma_C(R))$ = $\{t_2, t_5\}$ for $M = \mathcal{M}$, and $\lambda_{M}(\sigma_C(R))$ = $\{t_2\}$ for $M = \{m_1\}$.\qed\vspace{-2mm}
\end{example}

{\flushleft \textbf{Problem Statement}}\hspace{2mm}  Given an append-only table $R(\mathcal{D};\mathcal{M})$ and the last tuple $t$ that was appended onto $R$, the \emph{situational fact discovery problem}
is to find each constraint-measure pair $(C,M)$ such that $t$ is in the contextual skyline.  The result, denoted $S^t$, is $\{(C,M)| C$$\in$$\mathcal{C}_{\mathcal{D}}, M$$\subseteq$$\mathcal{M}, t$$\in$$\lambda_{M}(\sigma_C(R))\}$.  For simplicity of notation, we call $S^t$ ``the contextual skylines for $t$'', even though rigorously speaking it is the set of $(C,M)$ pairs whose corresponding contextual skylines include $t$.




\section{Solution Overview}\label{sec:idea}
\vspace{-1mm}
Discovering situational facts for a new tuple $t$ entails finding
constraint-measure pairs that qualify $t$ as a contextual skyline
tuple.  We identify three sources of inefficiency in
a straightforward brute-force method, and we propose
corresponding ideas to tackle them.  To facilitate the discussion,
we define the concept of \emph{tuple-satisfied constraints}, which are
all constraints pertinent to $t$, corresponding to the contexts
containing $t$.\vspace{-1.5mm}

\begin{definition}[Tuple-Satisfied Constraint]
Given a tuple $t \in R(\mathcal{D};\mathcal{M})$ and a constraint $C \in \mathcal{C}_{\mathcal{D}}$, if $\forall d_i \in \mathcal{D}$, $C.d_i = *$ or $C.d_i = t.d_i$, we say $t$ \emph{satisfies} $C$. We denote the set of all such satisfied constraints by $\mathcal{C}_{\mathcal{D}}^t$ or simply $\mathcal{C}^t$ when $\mathcal{D}$ is clear in context.  It follows that given any $C \in \mathcal{C}^t$, $t \in \sigma_C(R)$.\qed\vspace{-2mm}
\end{definition}

For $C$$\in$$\mathcal{C}^t$, $C.d_i$ can attain two possible values $\{*, t.d_i\}$.
Hence, $\mathcal{C}^t$ has $2^n$ constraints in total for $|\mathcal{D}|$$=$$n$.
Alg.\algfindconstraints\ is a simple routine used in all algorithms for finding all constraints of $\mathcal{C}^{t}$.
It generates the constraints from the most general constraint $\top$$=$$\langle *, *, \ldots, * \rangle$
to the most specific constraint $\langle t.d_1, t.d_2, \ldots, t.d_n \rangle$.
$\top$ has no bound attributes, i.e., $bound(\top)$$=$$0$.
Alg.\algfindconstraints\ makes sure a constraint is not generated twice, for efficiency, by not continuing the while-loop in Line~\linewhileloop\ once a specific attribute value is found in $C$.

A brute-force approach to the contextual skyline discovery problem
would compare a new tuple $t$ with every tuple in $R$ to determine if
$t$ is dominated, repeatedly for every constraint satisfied by $t$ in
every possible measure subspace.  It is shown in
Alg.\algbruteforce.  The obvious inefficiency of this approach
has three culprits---the exhaustive comparison with \emph{every tuple}, for
\emph{every constraint} and in \emph{every measure subspace}.  We
devise three corresponding ideas to counter these causes, as
follows:\vspace{-1.5mm}

{\flushleft \textbf{(1)}} \textbf{Tuple reduction}\hspace{2mm}
For a constraint-measure pair $(C,M)$, $t$ is in the
contextual skyline $\lambda_M(\sigma_C(R))$ if $t$ belongs to
$\sigma_C(R)$ and is not dominated by any tuple in
$\sigma_C(R)$.  Instead of comparing $t$ with every tuple, it suffices
to only compare with current skyline tuples.  This simple
optimization is based on the following proposition which ways, if any tuple
dominates $t$, there must exist a skyline tuple that also
dominates $t$. \vspace{-2mm}

\begin{proposition}
\label{prop:cmp_skyline}
\textit{Given a new tuple $t$ inserted into $R$, a constraint $C \in \mathcal{C}^t$
and a measure subspace $M$, $t \in \lambda_M(\sigma_C(R))$ if and only if
$\nexists\ t' \in \lambda_M(\sigma_C(R))$ such that $t' \succ_M t$.}\qed\vspace{-2mm}
\end{proposition}

To exploit this idea, our algorithms conceptually maintain the
contextual skyline tuples for each context (i.e., measure subspace and
constraint), and compare $t$ only with these tuples for constraints
that $t$ satisfies. \vspace{-2mm}

{\flushleft \textbf{(2)}} \textbf{Constraint pruning}\hspace{2mm}
For constraints satisfied by $t$, we need to determine whether
$t$ enters the contextual skyline.  To prune constraints
from consideration, we note the following property: if
$t$ is dominated by a skyline tuple $t'$ under measure subspace $M$,
$t$ is not in the contextual skyline of constraint-measure pair
$(C,M)$ for any $C$ satisfied by both $t$ and $t'$.

To enable constraint pruning, we organize all constraints
in $\mathcal{C}^t$ into a lattice by their subsumption relation.
The constraints satisfied by both $t$ and $t'$, denoted
$\mathcal{C}^{t,t'}$, also form a lattice, which is the intersection
of lattices $\mathcal{C}^{t}$ and $\mathcal{C}^{t'}$.  Below we
formalize the concepts of lattice and lattice
intersection.\vspace{-1mm}
\begin{definition}[Constraint Subsumption]\label{def:subsume}
Given $C_1, C_2 \in \mathcal{C}_{\mathcal{D}}$, $C_1$ is subsumed by or equal to $C_2$ (denoted $C_1 \trianglelefteq C_2$ or $C_2 \trianglerighteq C_1$) iff\vspace{-1mm}
  \begin{enumerate} \parskip0pt
    \item $\forall d_i \in \mathcal{D}$, $C_2.d_i = C_1.d_i$ or $C_2.d_i = *$.\vspace{-1mm}
  \end{enumerate}
$C_1$ is subsumed by $C_2$ (denoted $C_1 \vartriangleleft C_2$ or $C_2 \vartriangleright C_1$) iff $C_1 \trianglelefteq C_2$ but $C_1 \neq C_2$.  In other words, the following condition is also satisfied in addition to the above one---\vspace{-1mm}
  \begin{enumerate}
    \setcounter{enumi}{1}
    \item $\exists d_i \in \mathcal{D}$ such that $C_2.d_i$$=$$*$ and $C_1.d_i$$\neq$$*$, i.e, $d_i$ is bound to a value belonging to $dom(d_i)$ in $C_1$ but is unbound in $C_2$.\vspace{-1mm}
  \end{enumerate}
By definition, $\sigma_{C_1}(R) \subseteq \sigma_{C_2}(R)$ if $C_1 \trianglelefteq C_2$.\qed\vspace{-1mm}
\end{definition}

\begin{example}
Consider $C_1$$=$$\langle a, b, c \rangle$ and $C_2$$=$$\langle a, *, c \rangle$. Here $C_1.d_1$$=$$C_2.d_1$, $C_1.d_3$$=$$C_2.d_3$, $C_1.d_2$$=$$b$ and $C_2.d_2$$=$$*$. By Definition~\ref{def:subsume}, $C_1$ is subsumed by $C_2$, i.e. $C_1 \triangleleft C_2$.\qed\vspace{-1mm}
\end{example}

\begin{definition}[Partial Order on Constraints]
The subsumption relation $\trianglelefteq$ on $\mathcal{C}_{\mathcal{D}}$ forms a partial order.
The partially ordered set (poset) ($\mathcal{C}_{\mathcal{D}}$, $\trianglelefteq$) has
a \emph{top element} $\top = \langle *, *, \ldots, * \rangle$ that subsumes every other constraint in $\mathcal{C}_{\mathcal{D}}$.  $\top$ is the most general constraint, since it has no bound attributes.
Note that ($\mathcal{C}_{\mathcal{D}}$, $\trianglelefteq$) is not a lattice and does not have a single bottom element.
Instead, it has multiple \emph{minimal elements}.
Every minimal element $C$ satisfies the condition that $\forall d_i$, $C.d_i \neq *$.

If $C_1 \vartriangleleft C_2$, we say $C_1$ is a descendant of $C_2$ ($C_2$ is an ancestor of $C_1$).
If $C_1 \vartriangleleft C_2$ and $bound(C_1) - bound(C_2) = 1$, then $C_1$ is a child of $C_2$ ($C_2$ is a parent of $C_1$).
Given $C \in \mathcal{C}_{\mathcal{D}}$, we denote $C$'s ancestors, descendants, parents and children by $\mathcal{A}_C$,  $\mathcal{D}_C$, $\mathcal{P}_C$ and $\mathcal{CH}_C$, respectively.\qed\vspace{-1mm}
\end{definition}

\begin{definition}[Lattice of Tuple-Satisfied Constraints]
\label{def:latticeC}
Given $t$$\in$$R(\mathcal{D};\mathcal{M})$, $\mathcal{C}^t$$\subseteq$$\mathcal{C}_{\mathcal{D}}$ by definition.  In fact, ($\mathcal{C}^t, \trianglelefteq$) is a lattice.  Its top element is $\top$.  Its bottom element $\langle t.d_1, t.d_2, \ldots, t.d_n \rangle$, denoted $\bot(\mathcal{C}^t)$, is a minimal element in $\mathcal{C}_\mathcal{D}$.

Given $C$$\in$$\mathcal{C}^t$, we denote $C$'s ancestors, descendants, parents and children within $\mathcal{C}^t$ by $\mathcal{A}^t_C$,  $\mathcal{D}^t_C$, $\mathcal{P}^t_C$ and $\mathcal{CH}^t_C$, respectively.
$|\mathcal{CH}^t_C|$$=$$n$$-$$bound(C)$ where $n$$=$$|\mathcal{D}|$, i.e., each child of $C$ is a constraint by adding conjunct $d_i$$=$$t.d_i$ into $C$ for unbound attribute $d_i$.  It is clear that $|\mathcal{P}^t_C|$$=$$bound(C)$.
By definition, $\mathcal{A}^t_C$$=$$\mathcal{A}_C$ and $\mathcal{P}^t_C$$=$$\mathcal{P}_C$, while $\mathcal{D}^t_C$$\subseteq$$\mathcal{D}_C$ and $\mathcal{CH}^t_C$$\subseteq$$\mathcal{CH}_C$.\qed
\end{definition}

\begin{figure}[t]
\noindent \begin{minipage}{0.23\textwidth}
\centering
\epsfig{file=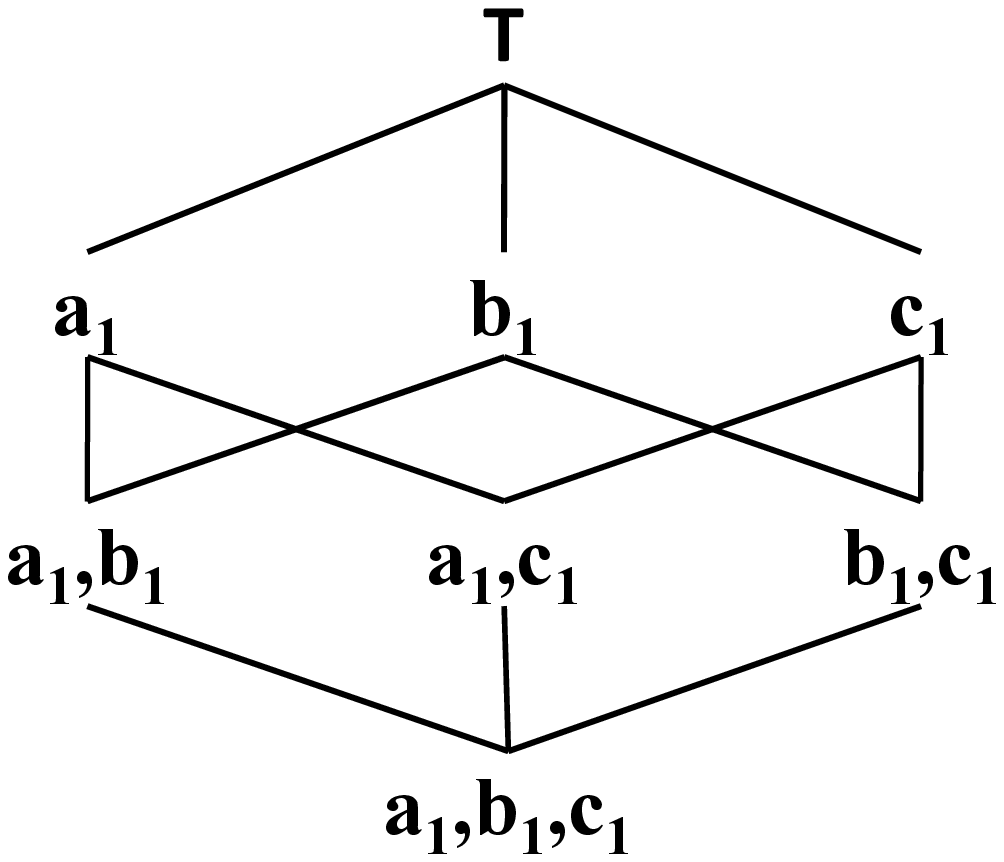,  scale = 0.32}\vspace{1mm}
\caption{\small Lattice $\mathcal{C}^{t_5}$}
\label{fig:cuboid}
\end{minipage}
\noindent \begin{minipage}{0.26\textwidth}
\centering
\epsfig{file=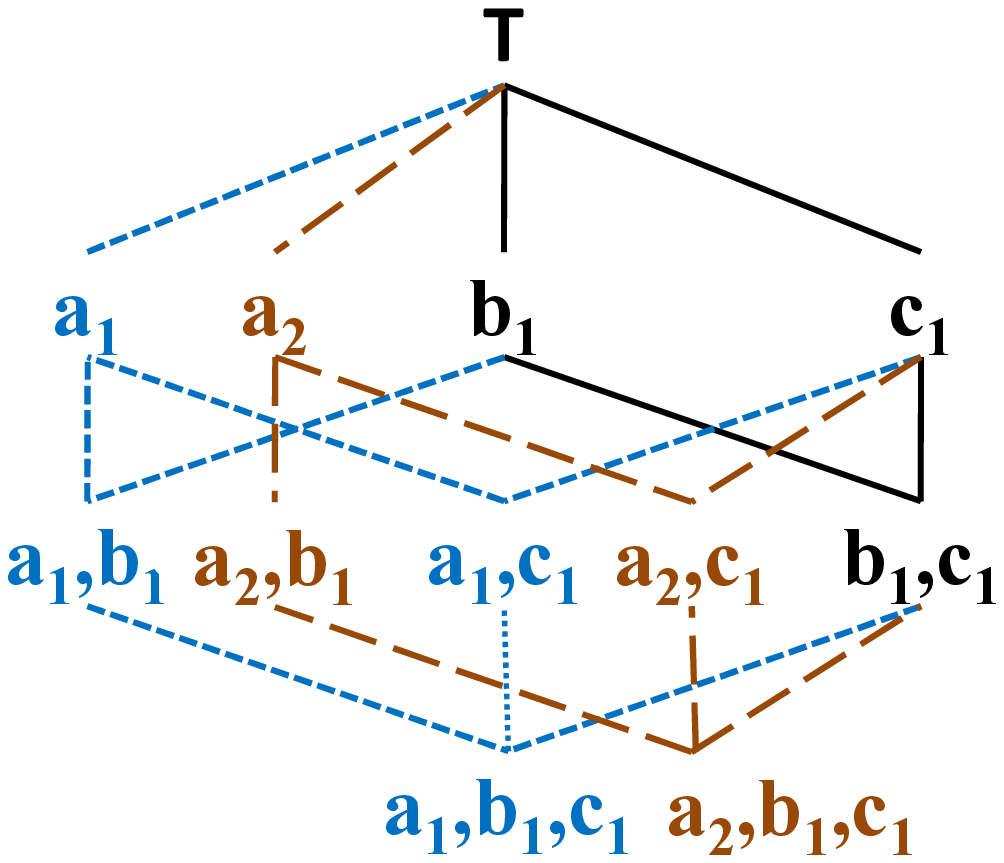, scale = 0.32}\vspace{1mm}
\caption{\small Intersection of $\mathcal{C}^{t_4}$ and $\mathcal{C}^{t_5}$}
\label{fig:sublattice}
\end{minipage}
\end{figure}

\begin{example}
Fig.\ref{fig:cuboid} presents lattice $\mathcal{C}^{t_5}$ for $t_5$ in Table~\ref{tab:running_example}.
For simplicity, we omit values on unbound dimension attributes (e.g., $\langle *, *, c_1 \rangle$ is represented as $c_1$).
Consider $C = \langle a_1, *, c_1 \rangle$.
$\mathcal{A}^{t_5}_C = \{\top, \langle a_1, *, * \rangle, \langle *, *, c_1 \rangle\}$, $\mathcal{P}^{t_5}_C = \{\langle a_1, *, * \rangle, \langle *, *, c_1 \rangle\}$, $\mathcal{CH}^{t_5}_C = \{\langle a_1, b_1, c_1 \rangle \}$ and $\mathcal{D}^{t_5}_C = \{\langle a_1, b_1, c_1 \rangle\}$.\qed
\end{example}

\begin{definition}[Lattice Intersection]~\label{def:latticeIntersect}
Given $t, t' \in R(\mathcal{D};\mathcal{M})$, $\mathcal{C}^{t,t'}$$=$$\mathcal{C}^{t} \cap \mathcal{C}^{t'}$ is the intersection of lattices $\mathcal{C}^{t}$ and $\mathcal{C}^{t'}$.
$\mathcal{C}^{t,t'}$ is non-empty and is also a lattice.
By Definition~\ref{def:latticeC}, the lattices for all tuples share the same top element $\top$.
Hence $\top$ is also the top element of $\mathcal{C}^{t,t'}$.
Its bottom $\bot(\mathcal{C}^{t,t'})$$=$$\langle v_1, v_2, \ldots, v_n \rangle$ where $v_i$$=$$t.d_i$ if $t.d_i$$=$$t'.d_i$ and $v_i$$=$$*$ otherwise.
$\bot(\mathcal{C}^{t,t'})$ equals $\top$ when $t$ and $t'$ do not have common attribute value.\qed
\end{definition}

\begin{example}
Fig.\ref{fig:sublattice} shows $\mathcal{C}^{t_4}$ and $\mathcal{C}^{t_5}$ for $t_4$ and $t_5$ in Table~\ref{tab:running_example}.  The constraints connected by solid lines represent the lattice intersection $\mathcal{C}^{t_4,t_5}$. Its bottom is $\bot(\mathcal{C}^{t_4,t_5})$ = $\langle *, b_1, c_1 \rangle$. In addition to $\mathcal{C}^{t_4,t_5}$, $\mathcal{C}^{t_4}$ and $\mathcal{C}^{t_5}$ further include the constraints connected by dashed and  dotted lines, respectively.\qed
\end{example}


The algorithms we are going to propose consider the constraints in
certain lattice order, compare $t$ with skyline tuples associated with
visited constraints, and use $t$'s dominating tuples to prune
unvisited constraints from consideration---thereby reducing cost.
This idea of lattice-based pruning of constraints is justified by
Propositions~\ref{prop:prune_ancestor} and~\ref{prop:prune_lattice}
below. 

\begin{proposition}
\label{prop:prune_ancestor}
\textit{Given a tuple $t$, if $t \notin \lambda_M(\sigma_C(R))$, then $t \notin \lambda_M(\sigma_{C'}$ $(R))$, for all $C' \in \mathcal{A}_C$.}\qed 
\end{proposition}

If $t \prec_M t'$, then $t \notin \lambda_M(\sigma_{\bot(\mathcal{C}^{t,t'})}(R))$.
Hence, according to Proposition~\ref{prop:prune_ancestor}, we have the following Proposition~\ref{prop:prune_lattice}. 

\begin{proposition}
\label{prop:prune_lattice}
\textit{Given two tuples $t$ and $t'$, if $t \prec_M t'$, then $t \notin \lambda_M(\sigma_{C}$ $(R))$, for all $C \in \mathcal{C}^{t,t'}$.}\qed 
\end{proposition}

{\flushleft \textbf{(3)}} \textbf{Sharing computation across measure
  subspaces}\hspace{2mm}
Given $t$, we need to consider not only all constraints satisfied by
$t$, but also all possible measure subspaces.  Sharing computation
across measure subspaces is challenging because of anti-monotonicity
of dominance relation---a skyline tuple under space $M$ may or may not
be a skyline tuple in another space $M'$, regardless of whether $M'$
is a superspace or subspace of $M$~\cite{PeiYLJELWTYZ06}.  We thus propose algorithms that
first traverse the lattice in the full measure space, during which a
frontier of constraints is formed for each measure subspace.  Top-down
(respectively, bottom-up) lattice traversal in a subspace commences from (respectively, stops at)
the corresponding frontier instead of the root, which in effect prunes
some top constraints.

{\flushleft \textbf{Two Baseline Algorithms}}\hspace{2mm}
We introduce two baseline algorithms \algname{BaselineSeq} (Alg.\algbaseline)
and \algname{BaselineIdx}. They are not as naive as the brute-force Alg.\algbruteforce.
Instead, they exploit Proposition~\ref{prop:prune_lattice} straightforwardly.
Upon $t$'s arrival, for each subspace $M$, they identify existing tuples $t'$ dominating $t$.
\algname{BaselineSeq} sequentially compares $t$ with every existing tuple.
$S$ is initialized to be $\mathcal{C}^t$ (Line~\lineinitS).
Whenever \algname{BaselineSeq} encounters a $t'$ that dominates $t$,
it removes constraints in $\mathcal{C}^{t,t'}$ from $S$ (Line~\linesubtract).
By Proposition~\ref{prop:prune_lattice}, $t$ is not in the contextual
skylines for those constraints.  After $t$ is compared with all
tuples, the constraints having $t$ in their skylines remain in $S$.
The same is independently repeated for every $M$.
The pseudo code of \algname{BaselineIdx} is similar to
Alg.\algbaseline\ and thus omitted.  Instead of comparing $t$
with all tuples, \algname{BaselineIdx} directly finds tuples
dominating $t$ by a one-sided range query $\bigwedge_{m_i \in M} (m_i$$\geq$$t.m_i)$
using a \emph{k}-d tree~\cite{Bentley1979} on full measure
space $\mathcal{M}$.


\section{Algorithms}\label{sec:alg}
This section starts with algorithms \algname{BottomUp} (Sec.~\ref{sec:alg:bottom-up}) and
\algname{TopDown} (Sec.~\ref{sec:alg:top-down}), which exploit the
ideas of tuple reduction and constraint pruning.  We then extend them
to enable sharing of computation across measure
subspaces (Sec.~\ref{Integrating_Dimension_and_Measure_Attributes}).

Based on the tuple-reduction idea
(Proposition~\ref{prop:cmp_skyline}), a new tuple $t$ should be
included into a contextual skyline if and only if $t$ is not dominated
by any current skyline tuple in the context.  Therefore,
\algname{BottomUp} and \algname{TopDown} store and maintain skyline
tuples for each constraint-measure pair $(C,M)$ and compare $t$ with
only the skyline tuples.  For clarity of discussion, we differentiate
between the contextual skyline ($\lambda_M(\sigma_C(R))$) and the
space for storing it ($\mu_{C,M}$), since tuples stored in $\mu_{C,M}$
do not always equal $\lambda_M(\sigma_C(R))$, by our algorithm design.

The algorithms traverse, for each measure subspace $M$, the
lattice of tuple-satisfied constraints $\mathcal{C}^t$ by certain
order.  When a constraint $C$ is visited, the algorithms
compare $t$ with the skyline tuples stored in $\mu_{C,M}$. If $t$ is
dominated by $t'$, then $t$ does not belong to the contextual skyline
of constraint-measure pair $(C,M)$.  Further, based on the
constraint-pruning idea (Proposition~\ref{prop:prune_lattice}), $t$
does not belong to the contextual skyline of $(C',M)$ for any $C'$
satisfied by both $t$ and $t'$ (i.e., $C' \in \mathcal{C}^{t,t'}$).
This property allows the algorithms to avoid
comparisons with skyline tuples associated with such constraints.

The algorithms differ by how skyline tuples are stored in
$\mu_{C,M}$. \algname{BottomUp} stores a tuple for every
constraint that qualifies it as a contextual skyline tuple, while
\algname{TopDown} only stores it for the topmost such constraints.  In
our ensuing discussion, we use invariants to formalize what must be
stored in $\mu_{C,M}$.  The algorithms also differ in the traversing
order of the constraints in $\mathcal{C}^t$. \algname{BottomUp} visits the constraints
bottom-up, while \algname{TopDown} makes the traversal
top-down.  Our discussion focuses on how the invariants are kept true
under the algorithms' different traversal orders and execution logics.
The algorithms present space-time tradeoffs.
\algname{TopDown} requires less space than \algname{BottomUp}
since it avoids storing duplicate skyline tuples as much as possible.
The saving in space comes at the cost of execution efficiency, due to
more complex operations in \algname{TopDown}.

Pei et al.~\cite{PeiYLJELWTYZ06} proposed bottom-up and top-down algorithms
to compute skycube.  However, their algorithms are for the lattice of
measure subspaces instead of constraints.

\subsection{Algorithm \textsf{BottomUp}}\label{sec:alg:bottom-up}

\begin{algorithm}[t]
\caption{\algname{BottomUp}}
\label{alg:bottomup}
\LinesNumbered
\footnotesize

\KwIn{$R(\mathcal{M},\mathcal{D})$: existing tuples; $t$: the new tuple}

\KwOut{$S^t$: the contextual skylines for $t$}

\BlankLine

$S^t \leftarrow \emptyset$;

\ForEach{$M \subseteq \mathcal{M}$}
{
  \lForEach{$C \in \mathcal{C}^t$}{$C.\mathit{pruned} \leftarrow \KwSty{false}$}

  $Q \gets \emptyset$;
  $Q.\mathit{enqueue}(\bot(\mathcal{C}^t))$;~\label{line:queueInitBU}\\
  \While{{\bf not} $Q.\mathit{empty}()$}
  {
	$C \leftarrow Q.\mathit{dequeue}()$;\\		
	$\mathit{dominated} \leftarrow \KwSty{false}$;\\
	\ForEach{$t' \in \mu_{C,M}$}{
	    \uIf{$t \prec_M t'$}	
	    {
	    	$\mathit{dominated} \leftarrow \KwSty{true}$;\\
	    	\ForEach{$C' \in \mathcal{A}^t_C$}{
                  $C'.\mathit{pruned} \leftarrow \KwSty{true}$;
                  \KwSty{break};\label{line:prune-skip}
                }
	    }
	    \lElseIf{$t' \prec_M t$}{
              $\mu_{C,M}.\mathit{delete}(t')$\label{line:remove}
            }
	}\label{line:compare}
	\If{{\bf not} $\mathit{dominated}$}{
        $S^t \leftarrow S^t \cup \{(C,M)\}$;~\label{line:fact}\\
		$\mu_{C,M}.\mathit{insert}(t)$;~\label{line:insert}\\
		\ForEach{$C' \in \mathcal{P}^t_C$}
		{
			\lIf{$(${\bf not} $Q.\mathit{contains}(C'))$ {\bf and} $(${\bf not} $C'.\mathit{pruned})$}{$Q.\mathit{enqueue}(C')$}
		}\label{line:enqueue}
	}
  }
}
$R \leftarrow R \cup \{t\}$;

\Return{$S^t$};
\end{algorithm}

\algname{BottomUp} (Alg.\ref{alg:bottomup}) stores a tuple for
every such constraint that qualifies it as a contextual skyline tuple.
Formally, Invariant~\ref{invar:bottomup} is guaranteed to hold before
and after the arrival of any tuple. 

\begin{invariant}
\label{invar:bottomup}
\textit{$\forall C \in \mathcal{C}_\mathcal{D}$ and $\forall M \subseteq
\mathcal{M}$, $\mu_{C,M}$ stores all skyline tuples
$\lambda_M(\sigma_C(R))$.}\qed
\end{invariant}

Upon the arrival of a new tuple $t$, for each measure subspace $M$,
\algname{BottomUp} traverses the constraints in $\mathcal{C}^t$ in a
bottom-up, breadth-first manner.  The traversal starts from
Line~\ref{line:queueInitBU} of Alg.\ref{alg:bottomup}, where the
bottom of $\mathcal{C}^t$ is inserted into a queue $Q$.  As long as
$Q$ is not empty, \algname{BottomUp} visits the next constraint $C$ from
the head of $Q$ and compares $t$ with current skyline tuples in
$\mu_{C,M}$ (Line~\ref{line:compare}).  Various actions are taken,
depending on comparison outcome.  \textbf{1)}~If $t$ is dominated
by any $t'$, the comparison with remaining tuples in
$\mu_{C,M}$ is skipped (Line~\ref{line:prune-skip}). The tuple $t$ is
disqualified from not only $C$ but also all constraints in
$\mathcal{C}^{t,t'}$, by Proposition~\ref{prop:prune_lattice}.
Because \algname{BottomUp} stores a tuple in all constraints that
qualify it as a contextual skyline tuple, and because it traverses
$\mathcal{C}^t$ bottom-up, the dominating tuple $t'$ must be
encountered at the bottom of $\mathcal{C}^{t,t'}$.  \algname{BottomUp} thus
skips the comparisons with all tuples stored for $C$'s ancestors
(Line~\ref{line:prune-skip}).  \textbf{2)}~If $t$ dominates $t'$, $t'$ is removed from $\mu_{C,M}$
(Line~\ref{line:remove}).  \textbf{3)}~If $t$ is not dominated by any
tuple in $\mu_{C,M}$, it is inserted into $\mu_{C,M}$
(Line~\ref{line:insert}) and $(C,M)$ corresponds to a contextual
skyline for $t$ (Line~\ref{line:fact}).  Further, each parent
constraint of $C$ that is not already pruned is inserted into $Q$, for
continuation of bottom-up traversal (Line~\ref{line:enqueue}).

Below we prove that Invariant~\ref{invar:bottomup} is satisfied by \algname{BottomUp} throughout its execution
over all tuples. 

{\flushleft \textbf{Proof of Invariant~\ref{invar:bottomup}}}\hspace{2mm}
We prove by induction
on the size of table $R$.  Invariant~\ref{invar:bottomup} is trivially
true when $R$ is empty.  If the invariant is true before the arrival
of $t$, i.e., $\mu_{C,M}$ stores all tuples in
$\lambda_M(\sigma_C(R))$, we prove that it remains true after the
arrival of $t$.  The proof entails showing that both insertions into and
deletions from $\mu_{C,M}$ are correct.

With regard to insertion, the only place where a tuple can be inserted
into $\mu_{C,M}$ is Line~\ref{line:insert} of \algname{BottomUp}, which
is reachable if and only if $t$ is not dominated by any tuple in
$\mu_{C,M}$ and $C$ belongs to $\mathcal{C}^t$.  This ensures that
$\mu_{C,M}$ stores $t$ if and only if $t \in \lambda_M(\sigma_C(R))$.
Further, it enures that no previous tuple is inserted into $\mu_{C,M}$
upon the arrival of $t$, which is correct since such a tuple was not
even in the skyline before.

With regard to deletion, the only place where a previous skyline tuple
$t'$ can be deleted from $\mu_{C,M}$ is Line~\ref{line:remove}, which
is reachable if and only if $t$ dominates $t'$ and $C$ is satisfied by
both tuples.  This ensures that $t'$ is removed from $\mu_{C,M}$ if
and only if $t'$ is not a skyline tuple anymore.

Hence, regardless of whether insertion/deletion takes place upon
$t$'s arrival, $\mu_{C,M}$ stores all tuples in
$\lambda_M(\sigma_C(R))$ afterwards.
\qedproof

\begin{figure}[t]
\begin{subfigure}[b]{.5\linewidth}
\centering
   \epsfig{file=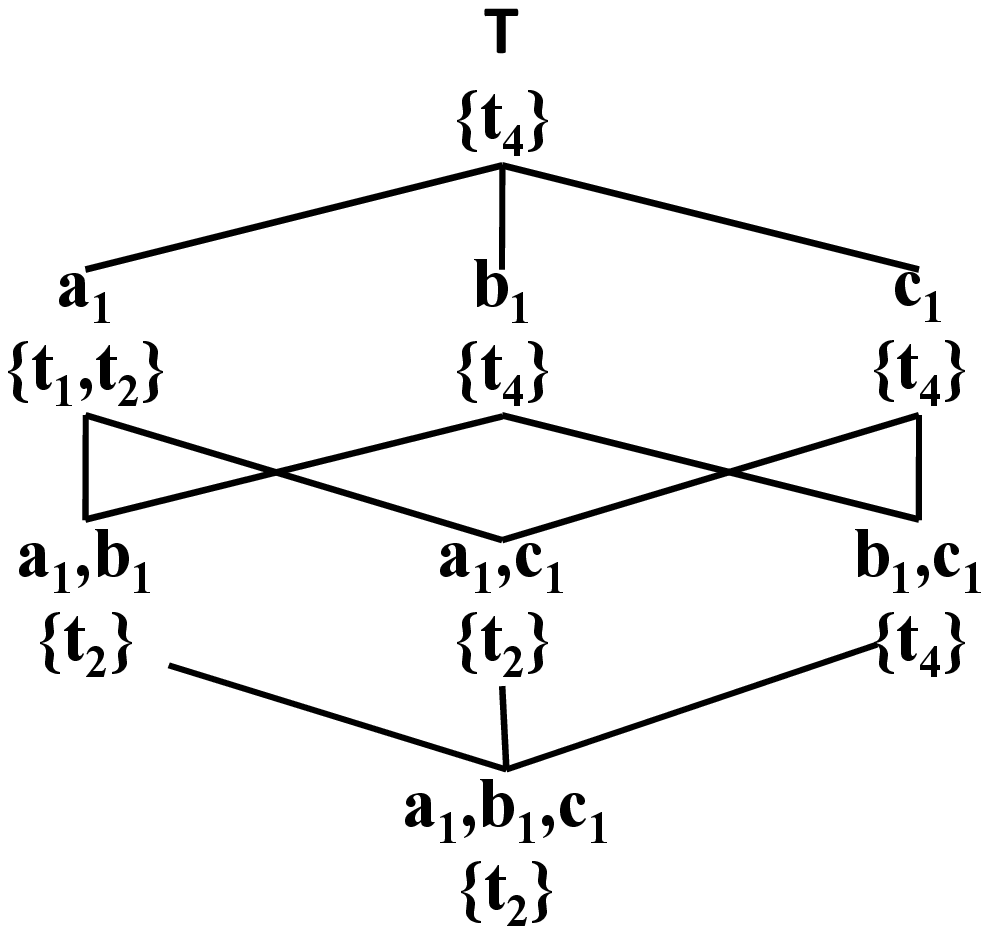, scale = 0.32}
   \caption{Before $t_5$}
   \label{fig:bottom_up_before}
\end{subfigure}
\hspace{-3mm}
\begin{subfigure}[b]{.5\linewidth}
\centering
   \epsfig{file=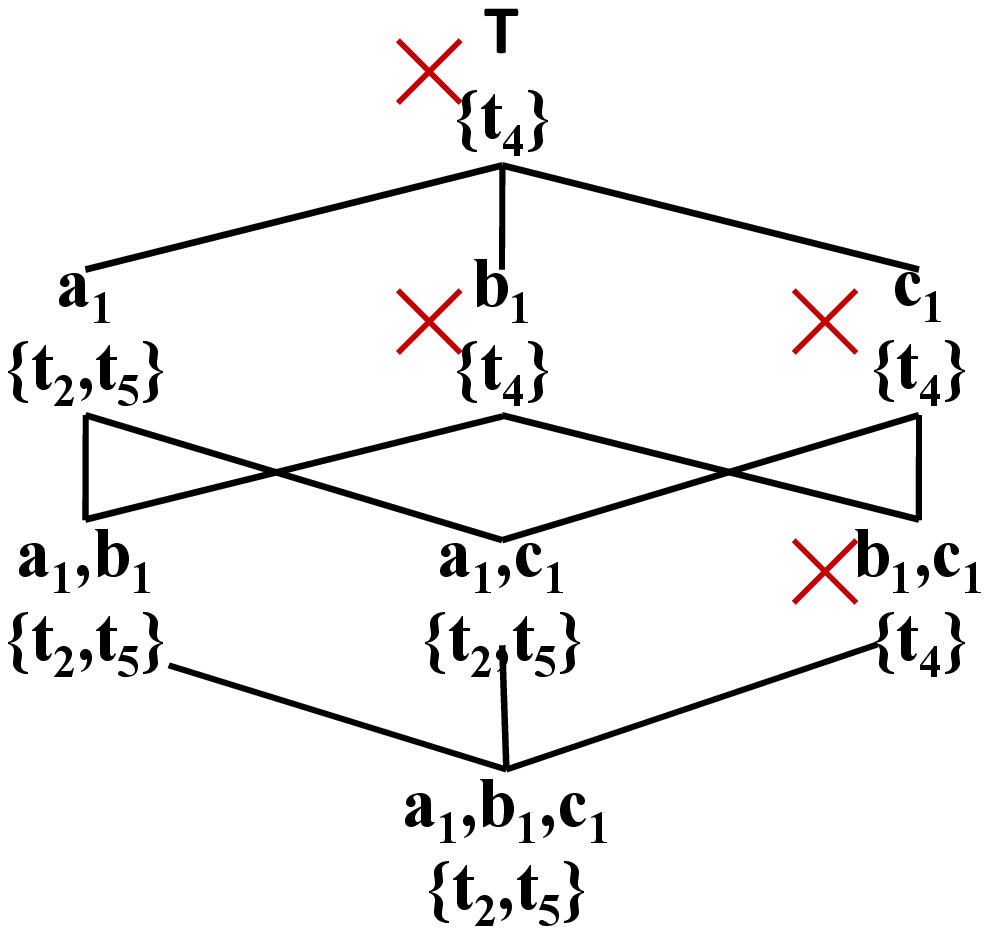, scale = 0.32}
   \caption{After $t_5$}
   \label{fig:bottom_up_after}
\end{subfigure}
   \vspace{-5mm}
 \caption{\small Execution of \algname{BottomUp} in Measure Subspace $\{m_1,m_2\}$} 
 \label{fig:bottom_up}
\end{figure}

\begin{example}
We use Fig.\ref{fig:bottom_up} to explain the execution of \algname{BottomUp} on Table~\ref{tab:running_example}, for measure subspace $M$=\{$m_1$,$m_2$\}.  Assume the tuples are inserted into the table in the order of $t_1$, $t_2$, $t_3$, $t_4$ and $t_5$.  Fig.\ref{fig:bottom_up_before} shows the lattice $\mathcal{C}^{t_5}$ before the arrival of $t_5$.  Beside each constraint $C$, the figure shows $\mu_{C,M}$. Upon the arrival of $t_5$, \algname{BottomUp} starts the traversal of $\mathcal{C}^{t_5}$ from its bottom $\bot(\mathcal{C}^{t_5})$=$\langle a_1, b_1, c_1 \rangle$.  There is one skyline tuple stored in $\mu_{\bot(\mathcal{C}^{t_5}),M}$---$t_2$.  In subspace $M$, $t_5$ is incomparable to $t_2$.  Hence, $t_5$ is inserted into it.  The traversal continues with the parents of $\bot(\mathcal{C}^{t_5})$.  Among its three parents, $\langle a_1, b_1, * \rangle$ and $\langle a_1, *, c_1 \rangle$ undergo the same insertion of $t_5$.   However, the contextual skyline for $\langle *, b_1, c_1 \rangle$ does not change, since $t_5$ is dominated by $t_4$ in $M$.  All constraints in $\mathcal{C}^{t_4,t_5}$ (i.e., $\langle *, b_1, c_1 \rangle$ and all its ancestors) are pruned from consideration by Property~\ref{prop:prune_lattice}.  The traversal continues at $\langle a_1, *, * \rangle$, for which $t_1$ is removed from the contextual skyline as it is dominated by $t_5$ in subspace $M$ and $t_5$ is inserted into it.  After that, the algorithm stops since there is no more unpruned constraints.  The content of $\mu_{C,M}$ for constraints in $\mathcal{C}^{t_5}$ after the arrival of $t_5$ is shown in Fig.\ref{fig:bottom_up_after}.\qed 
\end{example}


\subsection{Algorithm \textsf{TopDown}}\label{sec:alg:top-down}

\algname{BottomUp} stores $t$ for every constraint-measure pair that
qualifies $t$ as a contextual skyline tuple.  If
$t$ is stored in $\mu_{C,M}$, then $t$ is also stored in
$\mu_{C',M}$ for all $C'$$\in$$\mathcal{D}^t_C$, i.e., descendants of $C$
pertinent to $t$.  For this reason, \algname{BottomUp} repeatedly
compares a new tuple with a previous tuple multiple times.  Such
repetitive storage of tuples and comparisons increase both space
complexity and time complexity.  On the contrary,
\algname{TopDown} (Alg.\ref{alg:topdown2}) stores a tuple in
$\mu_{C,M}$ only if $C$ is a \emph{maximal skyline constraint} of the tuple, defined
as follows. 

\begin{definition}[Skyline Constraint]~\label{def:skylineConstraint}
Given $t \in R(\mathcal{D};\mathcal{M})$ and $M \subseteq \mathcal{M}$, the \emph{skyline constraints} of $t$ in $M$, denoted $\mathcal{SC}^t_M$, are the constraints whose contextual skylines include $t$.  Formally, $\mathcal{SC}^t_M = \{C| C \in \mathcal{C}^t, t \in \lambda_M(\sigma_C(R))\}$.  Correspondingly, other constraints in $\mathcal{C}^t$ are \emph{non-skyline constraints}.\qed 
\end{definition}

\begin{definition}[Maximal Skyline Constraints]~\label{def:msc} With
  regard to $t$ and $M$, a skyline constraint is a \emph{maximal
    skyline constraint} if it is not subsumed by any other skyline
  constraint of $t$.  The set of $t$'s maximal skyline constraints is
  denoted $\mathcal{MSC}^t_M$.  In other words, it includes those
  skyline constraints for which no parents (and hence ancestors) are
  skyline constraints. Formally, $\mathcal{MSC}^t_M = \{C| C \in
  \mathcal{SC}^t_M$, and $\nexists C' \in \mathcal{A}_C \text{ s.t. }
  C' \in \mathcal{SC}^t_M\}$.\qed 
\end{definition}

\begin{algorithm}[t]
\caption{\algname{TopDown}}
\label{alg:topdown2}
\SetInd{0.4em}{0.4em}
\LinesNumbered
\footnotesize
\KwIn{$R(\mathcal{M},\mathcal{D})$: existing tuples; $t$: the new tuple}

\KwOut{$S^t$: the contextual skylines for $t$}

\BlankLine

\columnsep=5pt
\begin{multicols}{2}

$S^t \leftarrow \emptyset$;

\ForEach{$M \subseteq \mathcal{M}$}
{
  \ForEach{$C \in \mathcal{C}^t$}
  {
    $C.\mathit{pruned} \leftarrow \KwSty{false}$;\\
    $C.\mathit{inAnces} \leftarrow \KwSty{false}$;
  }

  $Q \gets \emptyset$; $Q.\mathit{enqueue}(\top)$;~\label{line:queueInitTD}\\
  \While{{\bf not} $Q.\mathit{empty}()$}
  {
	$C \leftarrow Q.\mathit{dequeue}()$;\\
	\ForEach{$t' \in \mu_{C,M}$}
	{
	    \uIf{$t \prec_M t'$}	
	    {
			\algname{Dominated}$(t',C)$;\label{line:dominated}
	    }
	    \ElseIf{$t' \prec_M t$}
	    {
			\algname{Dominates}$(t',C,M)$;\label{line:removeTD}
	    }
	}\label{line:compareTD}
	\If{{\bf not} $C.\mathit{pruned}$}
	{
        $S^t \leftarrow S^t \cup \{(C,M)\}$;~\label{line:factTD}\\
        \If{{\bf not} $C.\mathit{inAnces}$}
	    {
    		$\mu_{C,M}.\mathit{insert}(t)$;~\label{line:insertTD}\\
    	}
	}
	\algname{EnqueueChindren}$(C)$;\label{line:enqueueTD}
  }
}
$R \leftarrow R \cup \{t\}$;

\Return{$S^t$};

\BlankLine
\BlankLine

\SetKwInput{KwProc}{Procedure}

\setcounter{AlgoLine}{0}
\SetKwFunction{dominates}{\algname{Dominates}}

\KwProc{\dominates$(t',C,M)$}

$\mu_{C,M}.\mathit{delete}(t')$;\\	
\ForEach{$C' \in \mathcal{CH}^{t'}_C-\mathcal{C}^{t}$}
{
	$\mathit{stored} \leftarrow \KwSty{false}$;\\
	\ForEach{$C'' \in \mathcal{A}^{t'}_{C'}-\mathcal{C}^{t}$}
	{
	 	\If{$t' \in \mu_{C'',M}$}
	    {
	    	$\mathit{stored} \leftarrow \KwSty{true}$;\\
	    	\KwSty{break};\\
	    }
	 }
	 \If{{\bf not} $\mathit{stored}$}
	 {
	 	$\mu_{C',M}.\mathit{insert}(t')$;~\label{line:pushdownTD}\\
	 }	    			
}

\hrule width 1.6in
\BlankLine
\setcounter{AlgoLine}{0}
\SetKwFunction{dominated}{\algname{Dominated}}

\KwProc{\dominated$(t',C)$}

$C.\mathit{pruned} \leftarrow \KwSty{true}$;

\ForEach{$C' \in \mathcal{C}^{t,t'}$}
{
	$C'.\mathit{pruned} \leftarrow \KwSty{true}$;
}

\hrule width 1.6in
\BlankLine
\setcounter{AlgoLine}{0}
\SetKwFunction{enqueue}{\algname{EnqueueChildren}}

\KwProc{\enqueue$(C)$}

\ForEach{$C' \in \mathcal{CH}^t_C$}
{
	\If{{\bf not} $C.\mathit{pruned}$}
	{
		$C'.\mathit{inAnces} \leftarrow \KwSty{true}$;		
	}
	\If{{\bf not} $Q.\mathit{contains}(C')$}
	{		
		$Q.\mathit{enqueue}(C')$;
	}
}	
\end{multicols}
\BlankLine
\end{algorithm}

\begin{figure}[t]
\begin{subfigure}[b]{.5\linewidth}
\centering
   \epsfig{file=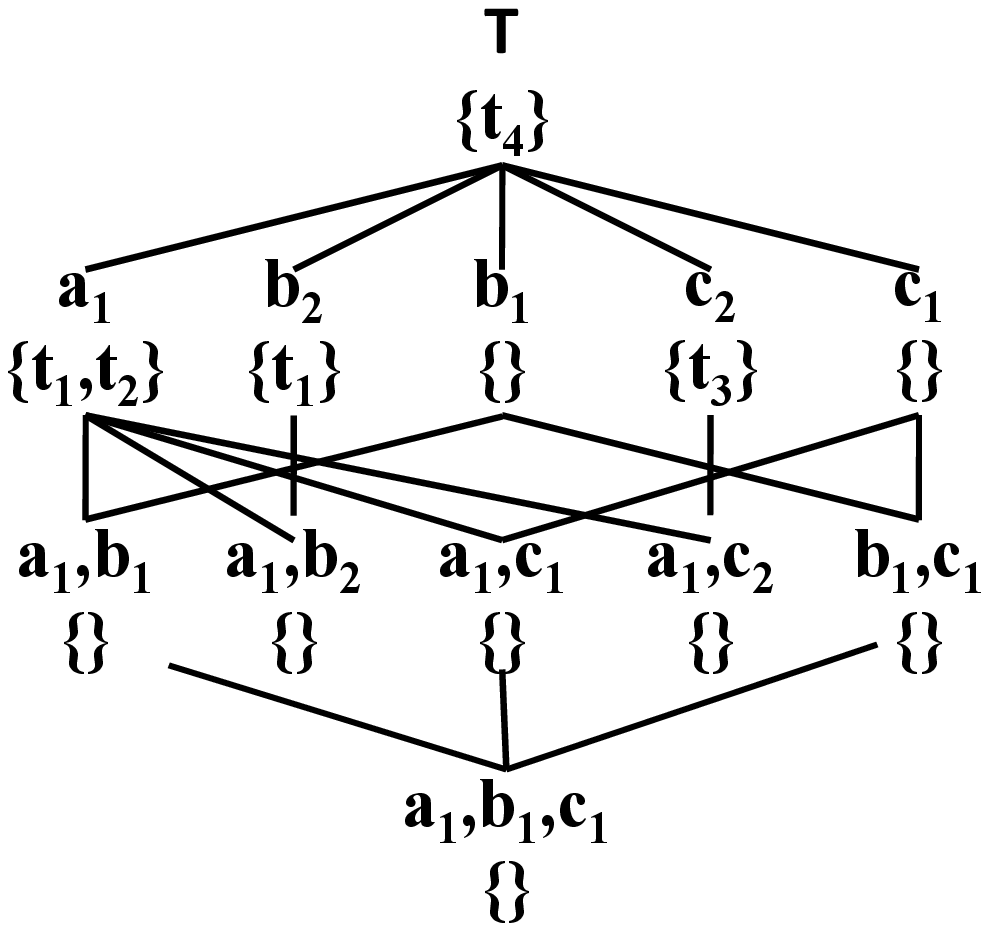, scale = 0.32}
   \caption{Before $t_5$}
   \label{fig:top_down_before}
\end{subfigure}
\hspace{-3mm}
\begin{subfigure}[b]{.5\linewidth}
\centering
   \epsfig{file=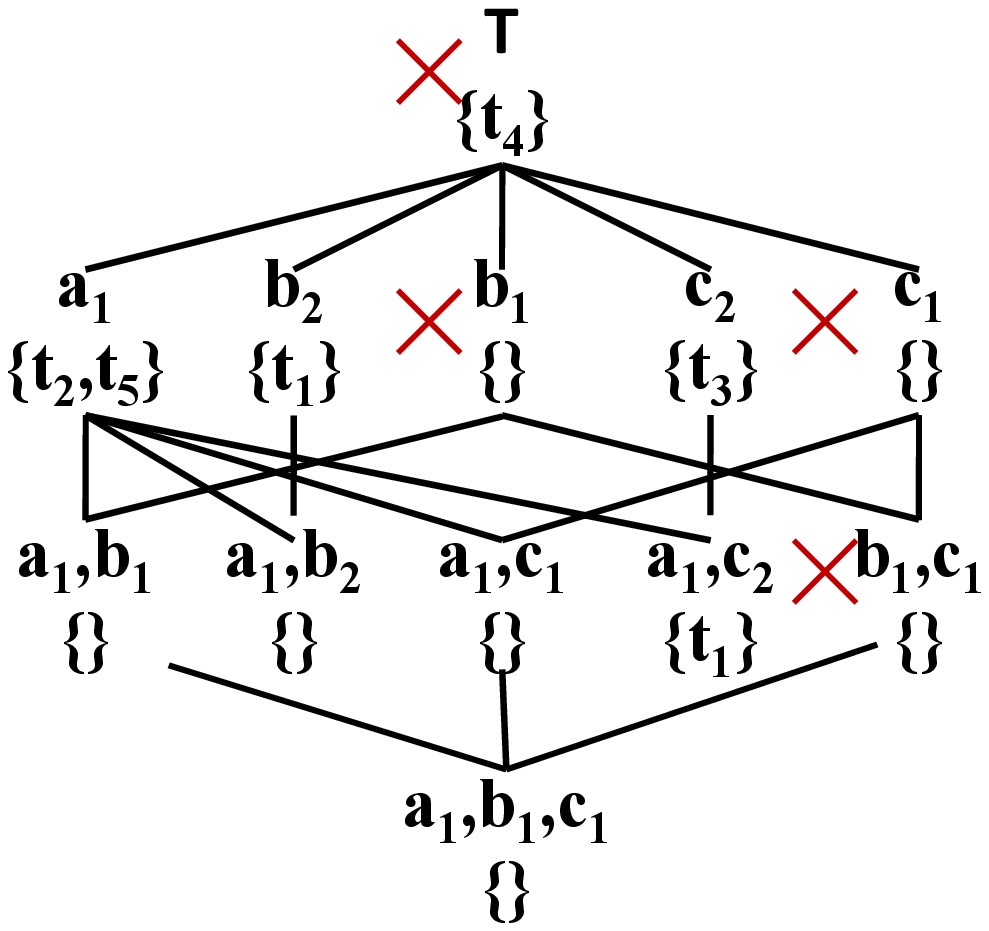, scale = 0.32}
   \caption{After $t_5$}
   \label{fig:top_down_after}
\end{subfigure}
   \vspace{-5mm}
 \caption{\small Execution of \algname{TopDown} in Measure Subspace $\{m_1,m_2\}$} 
 \label{fig:top_down}
\end{figure}

\begin{example}
Fig.\ref{fig:bottom_up_after} shows, in measure subspace $\{m_1$$,$$m_2\}$,
$t_5$ is in the contextual skylines of $4$ constraints, i.e.,
$\mathcal{SC}^{t_5}_{\{m_1,m_2\}}$ $=$ \{$\langle$$a_1$$,$$*,*$$\rangle$,$\langle$$a_1$$,$$b_1$$,$$*$$\rangle$,$\langle$$a_1$$,$$*$$,$$c_1$$\rangle$,$\langle$$a_1$$,$$b_1$$,$$c_1$$\rangle$\}. Its maximal skyline constraints are $\{\langle$$a_1$$,$$*$$,$$*$$\rangle\}$, i.e., $\mathcal{MSC}^{t_5}_{\{m_1,m_2\}}$$=$$\{$$\langle$$a_1$$,$$*$$,$$*$$\rangle$$\}$.\qed\vspace{-1mm}
\end{example}

Formally, Invariant~\ref{invar:topdown} is guaranteed by \algname{TopDown} before and after the arrival of any tuple. 

\begin{invariant}
\label{invar:topdown}
\textit{$\forall C \in \mathcal{C}_\mathcal{D}$ and $\forall M \subseteq \mathcal{M}$, $\mu_{C,M}$ stores a tuple $t$ if and only if $C \in \mathcal{MSC}^{t}_M$.}\qed
\end{invariant}

Different from \algname{BottomUp}, \algname{TopDown} stores a tuple in
its maximal skyline constraints $\mathcal{MSC}^{t}_M$ instead of all
skyline constraints $\mathcal{SC}^{t}_M$.  Due to this difference,
\algname{TopDown} traverses $\mathcal{C}^t$ in a
top-down (instead of bottom-up) breadth-first manner.  The traversal
starts from Line~\ref{line:queueInitTD} of
Alg.\ref{alg:topdown2}, where the top element $\top$ is inserted into a queue $Q$.  As long as $Q$ is not
empty, the algorithm visits the next constraint $C$ from the head of
$Q$ and compares $t$ with current skyline tuples in $\mu_{C,M}$
(Line~\ref{line:compareTD}).  Various actions are taken, depending on
the comparison result:

\textbf{1)} If $t$ is dominated by $t'$, $t$ is disqualified from not only $C$ but also all constraints in $\mathcal{C}^{t,t'}$, by Proposition~\ref{prop:prune_lattice}.  The pruning is done by calling \algname{Dominated} in Line~\ref{line:dominated} which sets $C'.pruned$ to true for every pruned constraint $C'$.  Since $C$ is a maximal skyline constraint for $t'$, the pruned constraints are all descendants of $C$ in $\mathcal{C}^{t,t'}$.  Note that \algname{TopDown} cannot skip the comparisons with the remaining tuples stored in $\mu_{C,M}$.  The reason is that there might be $t''$ in $\mu_{C,M}$ such that i)~$t''$ also dominates $t$ and ii)~$t''$ and $t$ share some dimension attribute values that are not shared by $t'$, i.e., $\mathcal{C}^{t,t''}$$-$$\mathcal{C}^{t,t'}$$\neq$$\emptyset$.  Since $t''$ is only stored in its maximal skyline constraints, skipping the comparison with $t''$ may incorrectly establish $t$ as a contextual skyline tuple for those constraints in $\mathcal{C}^{t,t''}$$-$$\mathcal{C}^{t,t'}$.

\textbf{2)} If $t$ dominates a current tuple $t'$, $t'$ is removed from $\mu_{C,M}$ by calling \algname{Dominates} (Line~\ref{line:removeTD}). An extra work is to update the maximal skyline constraints of $t'$ and store $t'$ in descendants of $C$ if necessary (Lines 2-9 of \algname{Dominates}). If $C$ has a child $C'$ satisfied by $t'$ but not $t$, $C'$ is a skyline constraint of $t'$. Further, $C'$ is a maximal skyline constraint of $t'$, if no ancestor of $C'$ is already a maximal skyline constraint of $t'$.

\textbf{3)}  If $t$ is not dominated by any tuple in $\mu_{C,M}$ and $C$ was not pruned before when its ancestors were visited, $(C,M)$ corresponds to a contextual skyline for $t$ (Line~\ref{line:factTD}).  If $t$ was not already stored in $C$'s ancestors (indicated by $C.\mathit{inAnces}$), then $C$ is a maximal skyline constraint and thus $t$ is inserted into $\mu_{C,M}$ (Line~\ref{line:insertTD}).

Furthermore, subroutine \algname{EnqueueChildren} is called for continuation of top-down traversal (Line~\ref{line:enqueueTD}).  It inserts each child constraint $C'$ of $C$ into $Q$.  If $t$ is stored in $\mu_{C,M}$ or any of its ancestors, $C'.\mathit{inAnces}$ is set to true and $t$ will not be stored again in $\mu(C',M)$ when the traversal reaches $C'$.

Below we prove that Invariant~\ref{invar:topdown} is satisfied by \algname{TopDown} throughout its execution over all tuples. 

{\flushleft \textbf{Proof of Invariant~\ref{invar:topdown}}}\hspace{2mm}
We prove by induction on the size of table $R$.  If the invariant is true before $t$'s arrival, i.e., $\mu_{C,M}$ stores a tuple $t$ if and only if $C$$\in$$\mathcal{MSC}^{t}_M$, we prove that it is kept true after the arrival of $t$.  The proof constitutes showing that both insertions into and deletions from $\mu_{C,M}$ are correct.

With regard to insertion, there are two places where a tuple can be inserted. \textbf{1)}~In Line~\ref{line:insertTD} of \algname{TopDown}, $t$ is inserted into $\mu_{C,M}$.  This line is reachable if and only if i)~$C$ is satisfied by $t$, ii)~$t$ is not dominated by any tuple stored at $C$ or $C$'s ancestors, and iii)~$t$ is not already stored at any of $C$'s ancestors.  This ensures that $\mu_{C,M}$ stores $t$ if and only if $C$ is a maximal skyline constraint of $t$, i.e., $C \in \mathcal{MSC}^{t}_M$.  \textbf{2)}~In Line~\ref{line:pushdownTD} of \algname{Dominates}, $t'$ is inserted into $\mu_{C',M}$.  This line is reachable if and only if i)~$t$ dominates $t'$, ii)~$C$, which is a parent of $C'$, is satisfied by both tuples, iii)~$C'$ is satisfied by $t'$ but not $t$, and iv)~$t'$ is not stored at any ancestor of $C'$.  Since $C$ was a maximal skyline constraint of $t'$ before the arrival of $t$, $C'$ must be a skyline constraint of $t$.  Therefore these conditions ensure that $\mu_{C',M}$ stores $t'$ if and only if $C'$ becomes a maximal skyline constraint of $t'$.

With regard to deletion, the only place where a previous skyline tuple $t'$ can be deleted from $\mu_{C,M}$ is Line~\ref{line:removeTD} of \algname{Dominates}, which is reachable if and only if $t$ dominates $t'$ and $C$ is satisfied by both tuples.  This ensures that $t'$ is removed from $\mu_{C,M}$ if and only if $C$ is not a maximal skyline constraint of $t'$ anymore.

Therefore, regardless of whether any insertion or deletion takes place upon the arrival of $t$, afterwards $\mu_{C,M}$ stores all tuples for which $C$ is a maximal skyline constraint.
\qedproof 

\begin{algorithm*}[t]
\caption{\algname{STopDown}}
\label{alg:subspace_sharing_distinct_value}
\SetInd{0.4em}{0.4em}
\LinesNumbered
\footnotesize

\columnsep=11pt
\begin{multicols}{3}
\KwIn{$R(\mathcal{M},\mathcal{D})$: existing tuples; $t$: the new tuple}

\KwOut{$S^t$: the contextual skylines for $t$}

\BlankLine
\vspace{2mm}

$S^t \leftarrow$ \algname{STopDownRoot}$()$;

\ForEach{$M \subset \mathcal{M}$}
{
	$S^t \leftarrow S^t \cup$ \algname{STopDownNode}$(M)$;
}

$R \leftarrow R \cup \{t\}$;

\Return{$S^t$};

\SetKwInput{KwProc}{Procedure}

\BlankLine
\hrule width 2in
\BlankLine
\setcounter{AlgoLine}{0}
\SetKwFunction{root}{\algname{STopDownRoot}}

\KwProc{\root$( )$}
\BlankLine

$S^t \leftarrow \emptyset$;

\ForEach{$C \in \mathcal{C}^t$}
{
  $C.\mathit{pruned} \leftarrow \KwSty{false}$;\\
  $C.\mathit{inAnces} \leftarrow \KwSty{false}$;\\
  \ForEach{$M \subset \mathcal{M}$}
  {
  	$\mathit{pruned}[C][M] \leftarrow \KwSty{false}$;
  }
}

$Q \gets \emptyset$; $Q.\mathit{enqueue}(\top)$;

\While{{\bf not} $Q.\mathit{empty}()$}
{
	$C \leftarrow Q.\mathit{dequeue}()$;\\
	\ForEach{$t' \in \mu_{C,\mathcal{M}}$}
	{
	    \lIf{$t \prec_{\mathcal{M}} t'$}{\algname{Dominated}$(t',C)$}

	    \lElseIf{$t' \prec_{\mathcal{M}} t$}{\algname{Dominates}$(t'$,$C$,$\mathcal{M})$}

	    \ForEach{$M \subset \mathcal{M}$}
	    {
	    	\If{$t \prec_{M} t'$ \emph{(Proposition~\ref{prop:subspace_sharing})}}
	    	{
	    		\ForEach{$C' \in \mathcal{C}^{t,t'}$}
	    		{
	    			$\mathit{pruned}[C'][M] \leftarrow \KwSty{true}$;\label{line:pruneSTD}
	    		}
	    	}
	    }\label{line:shareSTD}
	}\label{line:compareSTD}
	\If{{\bf not} $C.\mathit{pruned}$}
	{
        $S^t \leftarrow S^t \cup \{(C,\mathcal{M})\}$;\\
        \If{{\bf not} $C.\mathit{inAnces}$}
	    {
    		$\mu_{C,\mathcal{M}}.\mathit{insert}(t)$;\\
    	}
	}

	\algname{EnqueueChildren}$(C)$;
}
\Return{$S^t$};

\setcounter{AlgoLine}{0}
\SetKwFunction{node}{\algname{STopDownNode}}

\KwProc{\node$(M)$}
\BlankLine

$S^t \leftarrow \emptyset$;

\ForEach{$C \in \mathcal{C}^t$}
{
  $C.\mathit{pruned} \leftarrow \mathit{pruned}[C][M]$;\\
  $C.\mathit{inAnces} \leftarrow \KwSty{false}$;
}

$Q \gets \emptyset$; $Q.\mathit{enqueue}(\top)$;

\While{{\bf not} $Q.\mathit{empty}()$}
{
	$C \leftarrow Q.\mathit{dequeue}()$;	

	
	\If{{\bf not} $C.\mathit{pruned}$}
	{
		$S^t \leftarrow S^t \cup \{(C,M)\}$;\\
		\ForEach{$t' \in \mu_{C,M}$}
		{
			\lIf{$t' \prec_M t$}{\algname{Dominates}$(t',C,M)$\label{line:removeSTD}}
		}
		\If{{\bf not} $C.\mathit{inAnces}$}
		{
			$\mu_{C,M}.\mathit{insert}(t)$;~\label{line:insertSTD}\\
		}
	}	
	\algname{EnqueueChildren}$(C)$;
}
\Return{$S^t$};

\end{multicols}
\end{algorithm*}

\begin{example}
We use Fig.\ref{fig:top_down} to explain the execution of \algname{TopDown} on Table~\ref{tab:running_example} for $M$=\{$m_1$,$m_2$\}.  Again, assume the tuples are inserted into the table in the order of $t_1$, $t_2$, $t_3$, $t_4$ and $t_5$.  Fig.\ref{fig:top_down_before} shows $\mu_{C,M}$ beside each constraint $C$ in $\mathcal{C}^{t_5}$ before the arrival of $t_5$.  A tuple is only stored in its maximal skyline constraints.  The figure also shows constraints outside of $\mathcal{C}^{t_5}$ where various tuples are also stored.  The maximal skyline constraints for $t_2$ and $t_4$ are $\langle a_1, *, *\rangle$ and $\top$, respectively.  The maximal skyline constraints for $t_1$ include $\langle a_1, *, *\rangle$ and $\langle *, b_2, *\rangle$.  For $t_3$, the only maximal skyline constraint is $\langle *, *, c_2\rangle$.

Upon the arrival of $t_5$, \algname{TopDown} starts to traverse $\mathcal{C}^{t_5}$ from $\top$.  Only $t_4$ is stored in $\mu_{\top,M}$.  In $M$, $t_5$ is dominated by $t_4$, thus $\mu_{\top,M}$ does not change and $t_5$ does not belong to the contextual skylines of the constraints in $\mathcal{C}^{t_4, t_5}$---$\langle *, b_1, c_1 \rangle$, $\langle *, *, c_1 \rangle$, $\langle *, b_1, * \rangle$ and $\top$.  The traversal continues with the children of $\top$.  Among its three children, $\langle *, b_1, * \rangle$ and $\langle *, *, c_1 \rangle$ do not store any tuple, and $t_1$ and $t_2$ are stored at $\langle a_1, *, * \rangle$.  They do not dominate $t_5$ in $M$.  Since $t_5$ was not stored in any of its ancestors, $\langle a_1, *, * \rangle$ is a maximal skyline constraint of $t_5$.  Hence, $t_5$ is inserted into it and will not be stored at its descendants $\langle a_1, b_1, *\rangle$, $\langle a_1, *, c_1\rangle$ and $\langle a_1, b_1, c_1\rangle$.  Since $t_5$ dominates $t_1$, $t_1$ is deleted from $\langle a_1, *, * \rangle$. To update the maximal skyline constraints of $t_1$, \algname{TopDown} considers the two children of $\langle a_1, *, * \rangle$---$\langle a_1,b_2,*\rangle$ and $\langle a_1, *, c_2 \rangle$.  $\langle a_1, b_2, *\rangle$ is not a new maximal skyline constraint, since $t_1$ is already stored at its ancestor $\langle *,b_2,*\rangle$.  $\langle a_1,*,c_2\rangle$ becomes a new maximal skyline constraint since it is not subsumed by any existing maximal skyline constraint of $t_1$.  Thus $t_1$ is stored at $\langle a_1,*,c_2\rangle$. \algname{TopDown} continues to the end and finds no tuple at any remaining constraint in $\mathcal{C}^{t_5}$.  Fig.\ref{fig:top_down_after} depicts the content of $\mu_{C,M}$ for relevant constraints after $t_5$'s arrival.\qed
\end{example}

\subsection{Sharing across Measure Subspaces}
\label{Integrating_Dimension_and_Measure_Attributes}

Given a new tuple, both \algname{TopDown} and \algname{BottomUp}
compute its contextual skylines in each measure subspace separately,
without sharing computation across different subspaces.  As mentioned
in Sec.~\ref{sec:idea}, the challenge in such sharing lies
in the anti-monotonicity of dominance relation---with regard to the
same context of tuples, a skyline tuple in space $M$ may or may not
be a skyline tuple in another space $M'$, regardless of whether $M'$
is a superspace or subspace of $M$~\cite{PeiYLJELWTYZ06}.  To share computation across
different subspaces, we devise algorithms \algname{STopDown} and
\algname{SBottomUp}.  They discover the contextual skylines in all
subspaces by leveraging initial comparisons in the full measure space
$\mathcal{M}$.  In this section, we first introduce \algname{STopDown}
and then briefly explain \algname{SBottomUp}, which is based on
similar principles.

With regard to two tuples $t$ and $t'$, the measure space
$\mathcal{M}$ can be partitioned into three disjoint sets
$\mathcal{M}^{>}$, $\mathcal{M}^{<}$ and $\mathcal{M}^{=}$ such that
1)~$\forall m$$\in$$\mathcal{M}^{>}$, $t.m$$>$$t'.m$; 2)~$\forall
m$$\in$$\mathcal{M}^{<}$, $t.m$$<$$t'.m$; and 3)~$\forall
m$$\in$$\mathcal{M}^{=}$, $t.m$$=$$t'.m$.  Then, $t$ is dominated by
$t'$ in a subspace $M$ if and only if $M$ contains at least
one attribute in $\mathcal{M}^{<}$ and no attribute in
$\mathcal{M}^{>}$, as stated by
Proposition~\ref{prop:subspace_sharing}. 

\begin{proposition}
\label{prop:subspace_sharing}
\textit{In a measure subspace $M \subseteq \mathcal{M}$, $t \prec_M t'$ if and only if $M \cap \mathcal{M}^{<} \neq \emptyset$ and $M \cap \mathcal{M}^{>} = \emptyset$.}\qed
\end{proposition}

The gist of \algname{STopDown}
(Alg.\ref{alg:subspace_sharing_distinct_value}) is to compare a
new tuple $t$ with current tuples $t'$ in full
space $\mathcal{M}$ and, using
Proposition~\ref{prop:subspace_sharing}, identify all subspaces $M$ in
which $t'$ dominates $t$.  It starts by finding the skyline
constraints in $\mathcal{M}$ using
\algname{STopDownRoot}, which is similar to \algname{TopDown}
except Lines~\ref{line:shareSTD}-\ref{line:pruneSTD}.  While
traversing a constraint $C$, $t$ is compared with the tuples in
$\mu_{C,\mathcal{M}}$ (Line~\ref{line:compareSTD} of
\algname{STopDownRoot}).  By Proposition~\ref{prop:subspace_sharing},
all subspaces $M$ where $t'$ dominates $t$ are identified.
In each such $M$, constraints in
$\mathcal{C}^{t,t'}$ are pruned
(Lines~\ref{line:shareSTD}-\ref{line:pruneSTD})---indicated by setting
values in a two-dimensional matrix $\mathit{pruned}$.  After finishing
\algname{STopDownRoot}, for each $M$, the constraints $C$ satisfying
$\mathit{pruned}[C][M] = \KwSty{false}$ are
the skyline constraints of $t$ in $M$.  \algname{STopDown} then
continues to traverse these skyline constraints in $M$ by calling
\algname{STopDownNode}$(M)$, for two purposes---one is to store $t$ at
its maximal skyline constraints (Line~\ref{line:insertSTD} of
\algname{STopDownNode}), the other is to remove tuples dominated by
$t$ and update their maximal skyline constraints
(Line~\ref{line:removeSTD}). 

\begin{figure}[t]
\begin{subfigure}[b]{.5\linewidth}
\centering
   \epsfig{file=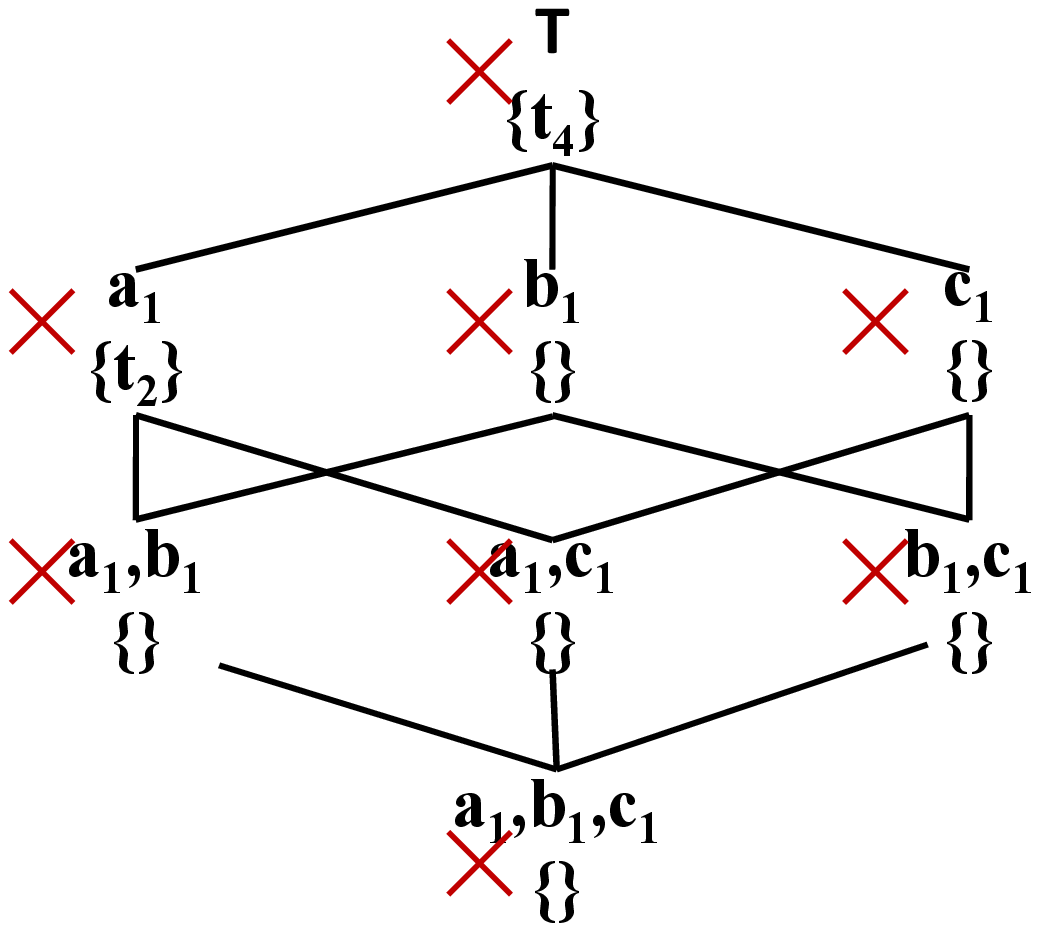, scale = 0.32}
   \caption{Before Visiting $\mathcal{C}^{t_5}$ in $\{m_1\}$}
   \label{fig:top_down_sharing_before_m1}
\end{subfigure}
\hspace{-2mm}
\begin{subfigure}[b]{.5\linewidth}
\centering
   \epsfig{file=./images/top_down_sharing_m1.eps, scale = 0.32}
   \caption{After Visiting $\mathcal{C}^{t_5}$ in $\{m_1\}$}
   \label{fig:top_down_sharing_after_m1}
\end{subfigure}
\caption{\small Execution of \algname{STopDown} in Measure Subspace $\{m_1\}$ (No Comparison Required and No Change Made)}
\label{fig:top_down_sharing_m1}
\end{figure}

\begin{figure}[t]
\begin{subfigure}[b]{.5\linewidth}
\centering
   \epsfig{file=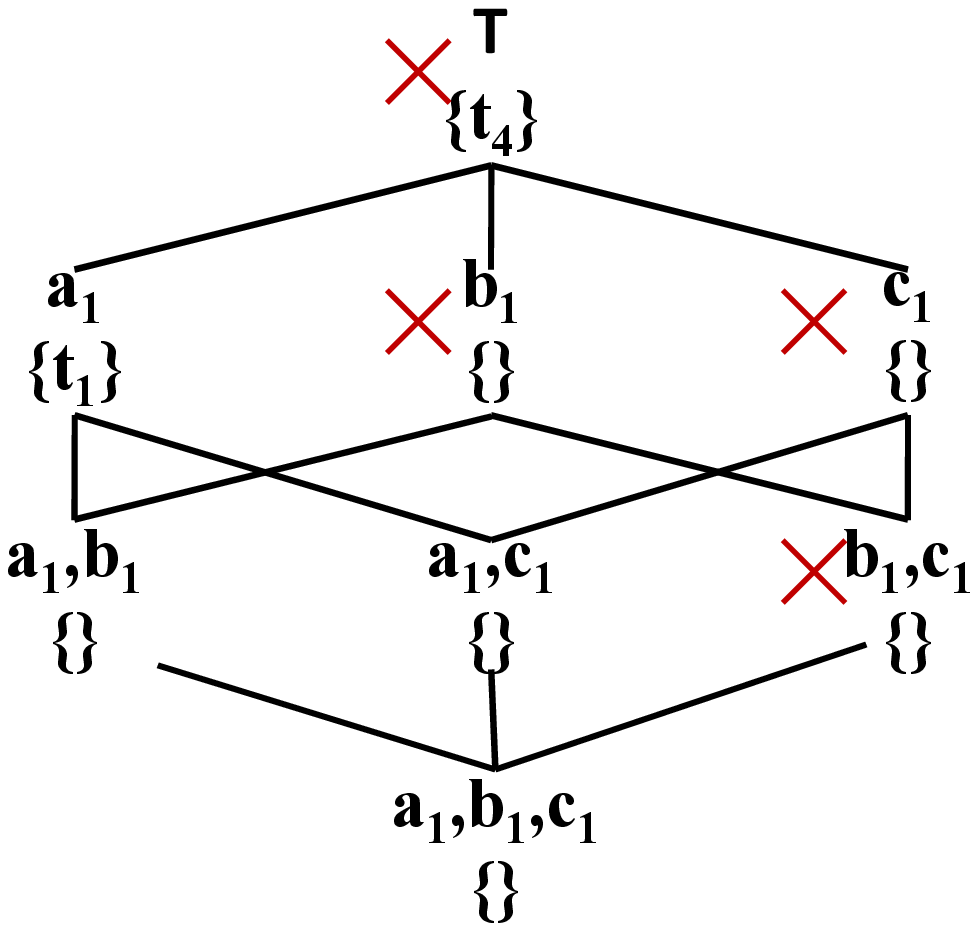, scale = 0.32}
   \caption{Before Visiting $\mathcal{C}^{t_5}$ in $\{m_2\}$}
   \label{fig:top_down_sharing_before_m2}
\end{subfigure}
\hspace{-2mm}
\begin{subfigure}[b]{.5\linewidth}
\centering
   \epsfig{file=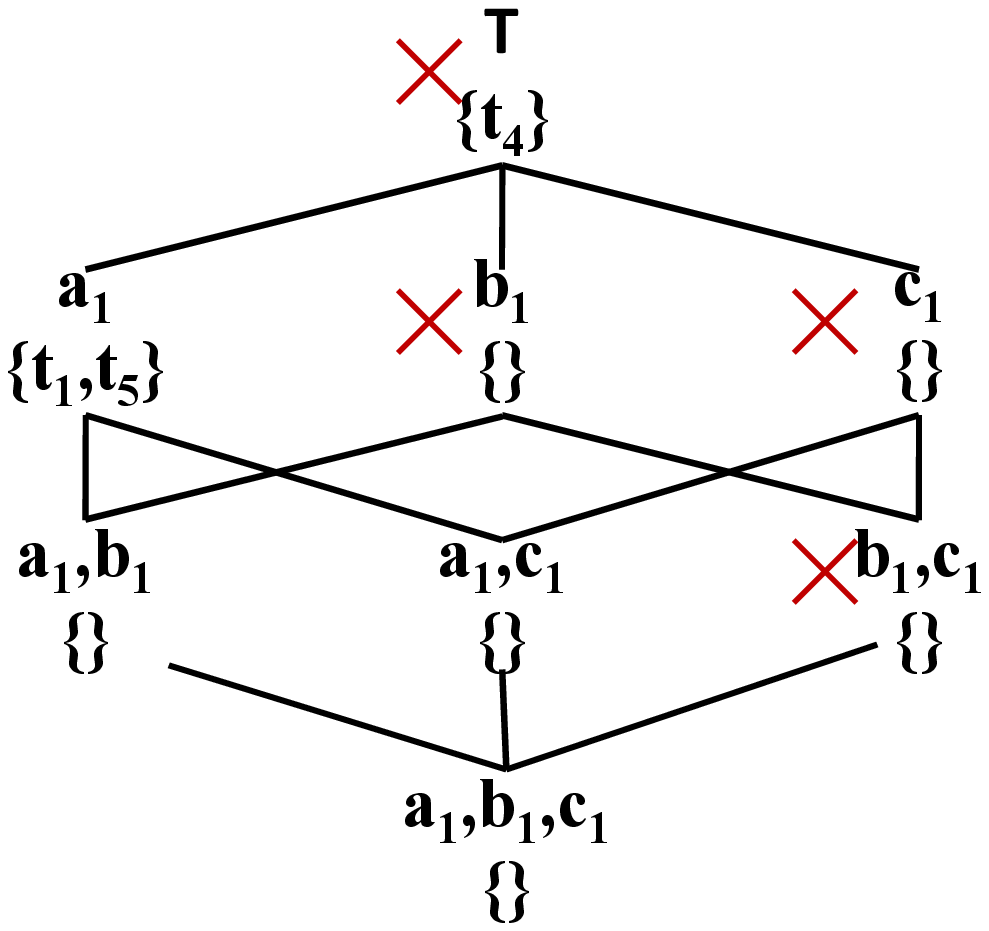, scale = 0.32}
   \caption{After Visiting $\mathcal{C}^{t_5}$ in $\{m_2\}$}
   \label{fig:top_down_sharing_after_m2}
\end{subfigure}
\caption{\small Execution of \algname{STopDown} in Measure Subspace $\{m_2\}$}
\label{fig:top_down_sharing_m2}
\end{figure}

\begin{example}
We explain \algname{STopDown}'s execution on Table~\ref{tab:running_example}.  In full space $\mathcal{M}$=$\{m_1,m_2\}$, \algname{STopDown} and \algname{TopDown} work the same.  Hence, Fig.\ref{fig:top_down} shows $\mu_{C,\mathcal{M}}$ beside each $C$ in $\mathcal{C}^{t_5}$ before and after $t_5$ arrives.  Comparisons with tuples in $\mathcal{M}$ also help to prune constraints in subspaces.  Consider $\top$ in Fig.\ref{fig:top_down_before}, where $t_4$ is stored.  The new tuple $t_5$ is compared with $t_4$.  The outcome is $\mathcal{M}^{>}$=$\emptyset$, $\mathcal{M}^{<}$=$\{m_1, m_2\}$ and $\mathcal{M}^{=}$=$\emptyset$, since $t_5$ is smaller than $t_4$ on both $m_1$ and $m_2$.  By Proposition~\ref{prop:subspace_sharing},  $t_5$ is dominated by $t_4$ in subspaces $\{m_1\}$ and $\{m_2\}$.   Hence, all constraints in $\mathcal{C}^{t_4,t_5}$ (including $\langle *, b_1, c_1 \rangle$, $\langle *, b_1, * \rangle$, $\langle *, *, c_1 \rangle$ and $\top$) are pruned in $\{m_1\}$ and $\{m_2\}$ simultaneously, by Lines~\ref{line:shareSTD}-\ref{line:pruneSTD} of \algname{STopDownRoot}.   As \algname{STopDownRoot} proceeds, $t_5$ is also compared with $t_1$ and $t_2$.  With regard to the comparison with $t_1$, since $\mathcal{M}^{<}$=$\emptyset$, $t_5$ is not dominated by $t_1$ in any space.  With regard to $t_2$, $\mathcal{M}^{>}$=$\{m_2\}$, $\mathcal{M}^{<}$=$\{m_1\}$ and $\mathcal{M}^{=}$=$\emptyset$.  Thus $t_5$ is dominated by $t_2$ in $\{m_1\}$.  Hence, all the constraints in $\mathcal{C}^{t_2,t_5}$, which is identical to $\mathcal{C}^{t_5}$, are pruned in $\{m_1\}$.

After the traversal in $\mathcal{M}$, \algname{STopDown} continues with each measure subspace.  In $\{m_1\}$, all constraints of $\mathcal{C}^{t_5}$ are pruned.  Hence, $t_5$ has no skyline constraint and nothing further needs to be done.  Fig.\ref{fig:top_down_sharing_m1} depicts $\mu_{C,\{m_1\}}$ for all $C$ in $\mathcal{C}^{t_5}$ before and after the arrival of $t_5$.  For $\{m_2\}$, Fig.\ref{fig:top_down_sharing_before_m2} depicts $\mu_{C,\{m_2\}}$ for all $C$ in $\mathcal{C}^{t_5}$ before the arrival of $t_5$.  Based on the analysis above, the skyline constraints of $t_5$ in $\{m_2\}$ include $\langle a_1, *, * \rangle$, $\langle a_1, b_1, * \rangle$, $\langle a_1, *, c_1 \rangle$ and $\langle a_1, b_1, c_1 \rangle$.   Since non-skyline constraints are pruned, $t_5$ is not compared with the tuples stored at those constraints.  Instead, $t_5$ is compared with $t_1$ stored at $\langle a_1, *, * \rangle$.  Since they do not dominate each other in $\{m_2\}$, $\langle a_1, *, * \rangle$ is a maximal skyline constraint of $t_5$ and $t_5$ is stored at it together with $t_1$.  The content of $\mu_{C,\{m_2\}}$ in $\mathcal{C}^{t_5}$ after encountering $t_5$ is in Fig.\ref{fig:top_down_sharing_after_m2}.  Note that \algname{TopDown} would have compared $t_5$ with other tuples seven times, including comparisons with $t_1$, $t_2$ and $t_5$ in $\{m_1, m_2\}$, with $t_2$ and $t_4$ in $\{m_1\}$, and with $t_1$ and $t_4$ in $\{m_2\}$.  In contrast, \algname{STopDown} needs four comparisons, including the same three comparisons in $\{m_1, m_2\}$ and another comparison with $t_1$ in $\{m_2\}$.\qed
\end{example}

Invariant~\ref{invar:topdown} is also guaranteed by \algname{STopDown}
all the time.  We omit the proof which is
largely the same as the proof for \algname{TopDown}.  We note the
essential difference between \algname{STopDown} and \algname{TopDown}
is the skipping of non-skyline constraints in measure
subspaces.  Since the new tuple is dominated under these constraints,
it does not and should not make any change to $\mu_{C,M}$ for any such
constraint-measure pair.

\algname{BottomUp} is extended to \algname{SBottomUp}, similar to how \algname{STopDown} extends \algname{TopDown}.  While in \algname{STopDown} lattice traversal in a measure subspace commences from the topmost skyline constraints instead of the root of a lattice, lattice traversal in \algname{SBottomUp} stops at them.  Invariant~\ref{invar:bottomup} is also warranted by \algname{SBottomUp}.  Its proof is similar to that for \algname{BottomUp}.  Due to space limitations, we do not further discuss \algname{SBottomUp}.

\section{Experiments}\label{sec:exp}
The algorithms were implemented in Java. The experiments were
conducted on a computer with $2.0$ GHz Quad Core 2 Duo Xeon CPU
running Ubontu 8.10. The limit on the heap size of Java
Virtual Machine (JVM) was set to $16$ GB. 

\subsection{Experiment Setup}

\textbf{Datasets}\hspace{2mm}
We used two real datasets, which exhibit similar trends.  We mainly discuss the results on
the NBA dataset.

\emph{NBA Dataset}\hspace{2mm} We collected $317{,}371$ tuples of NBA box scores from 1991-2004 regular seasons. We considered $8$ dimension attributes: \attrname{player}, \attrname{position}, \attrname{college}, \attrname{state}, \attrname{season}, \attrname{month}, \attrname{team} and \attrname{opp\_team}. \attrname{College} denotes from where a player graduated, if applicable. \attrname{State} records the player's state of birth. For measure attributes, $7$ performance statistics were considered: \attrname{points}, \attrname{rebounds}, \attrname{assists}, \attrname{blocks}, \attrname{steals}, \attrname{fouls} and \attrname{turnovers}. Smaller values are preferred on \attrname{turnovers} and \attrname{fouls}, while larger values are preferred on all other attributes.

\emph{Weather Dataset} ({\small \url{http://data.gov.uk/metoffice-data-archive}})\hspace{2mm} It has more than $7.8$ million daily weather forecast records collected from $5{,}365$ locations in six countries and regions of UK from Dec. 2011 to Nov. 2012. Each record has $7$ dimension attributes: \attrname{location}, \attrname{country}, \attrname{month}, \attrname{time step}, \attrname{wind direction [day]}, \attrname{wind direction [night]} and \attrname{visibility range} and $7$ measure attributes: \attrname{wind speed [day]}, \attrname{wind speed [night]}, \attrname{temperature [day]}, \attrname{temperature [night]}, \attrname{humidity [day]}, \attrname{humidity [night]} and \attrname{wind gust}.  We assumed larger values dominate smaller values on all attributes. 

\begin{figure*}[t]
\centering
\begin{subfigure}[b]{.33\textwidth}
\centering
   \epsfig{file=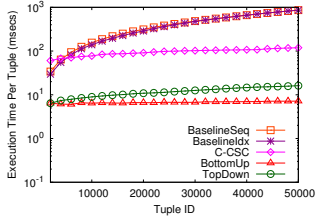, angle=-90, width=42mm,clip=}
   \caption{Varying $n$, $d$=$5$, $m$=$7$}
   \label{fig:3_3_-1_-1_MI_BI_BU_TD}
\end{subfigure}
\hspace{-2mm}
\begin{subfigure}[b]{.33\textwidth}
\centering
   \epsfig{file=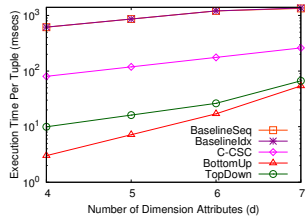, angle=-90, width=42mm,clip=}
   \caption{Varying $d$, $n$=$50{,}000$, $m$=$7$}
   \label{fig:1000_m5}
\end{subfigure}
\hspace{-2mm}
\begin{subfigure}[b]{.33\textwidth}
\centering
   \epsfig{file=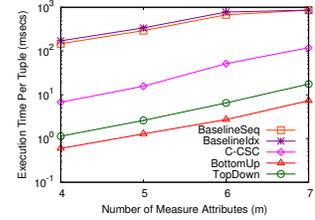, angle=-90, width=42mm,clip=}
   \caption{Varying $m$, $n$=$50{,}000$, $d$=$5$}
   \label{fig:1000_d5}
\end{subfigure}
\vspace{-2mm}
\caption{\small Execution Time of \algname{BaselineSeq}, \algname{BaselineIdx}, \algname{C-CSC}, \algname{BottomUp} and \algname{TopDown} on the NBA Dataset} 
\label{fig:BaselineBUTD}
\end{figure*}

\begin{figure*}[t]
\hspace{-5mm}
\noindent \begin{minipage}{0.75\textwidth}
\centering
\begin{subfigure}[b]{.3\textwidth}
   \epsfig{file=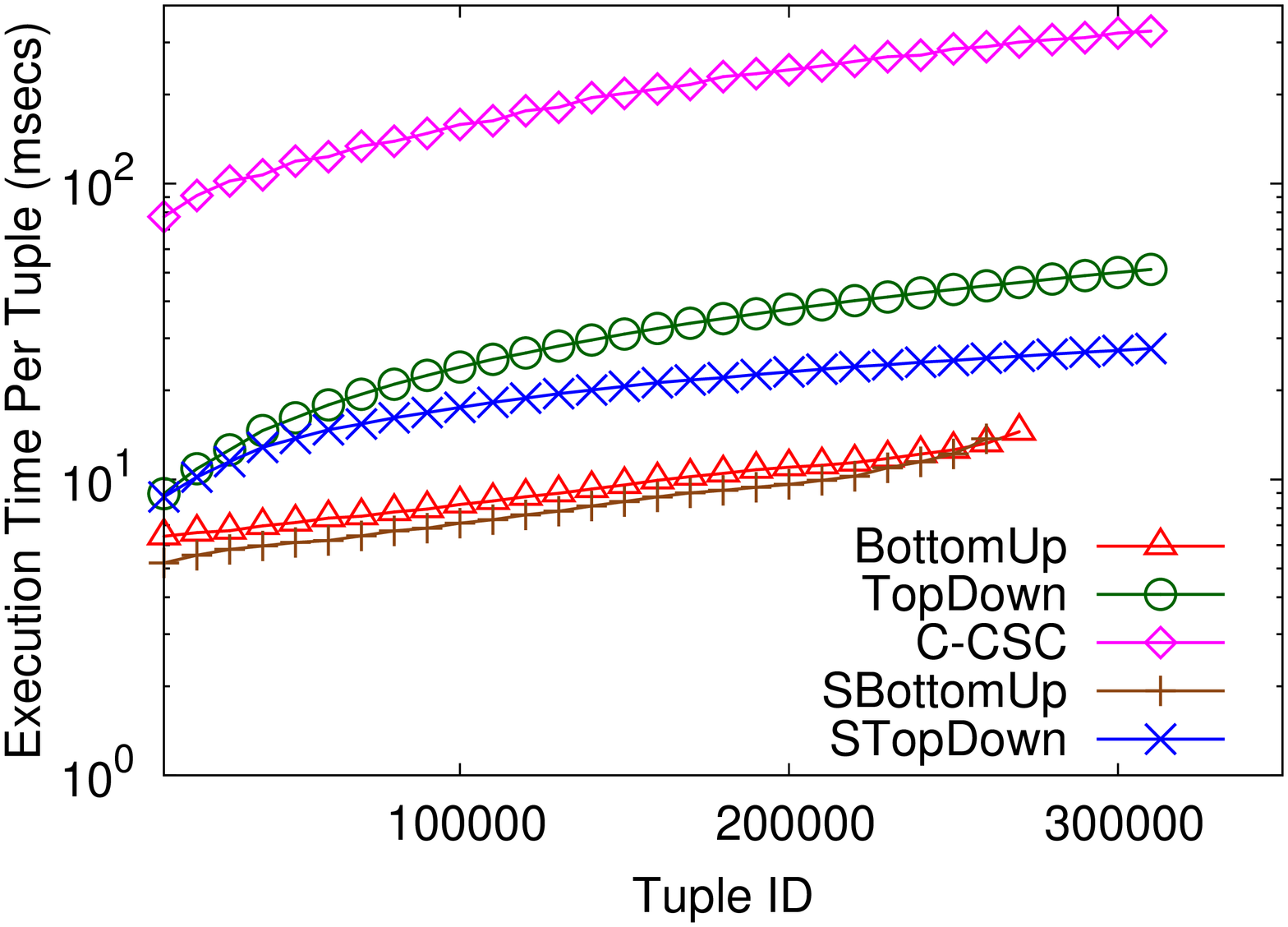,angle=-90, width=42mm,clip=}
   \caption{Varying $n$, $d$=$5$, $m$=$7$}
   \label{fig:8_7_300000_3_-1}
\end{subfigure}
\hspace{1mm}
\begin{subfigure}[b]{.3\textwidth}
   \epsfig{file=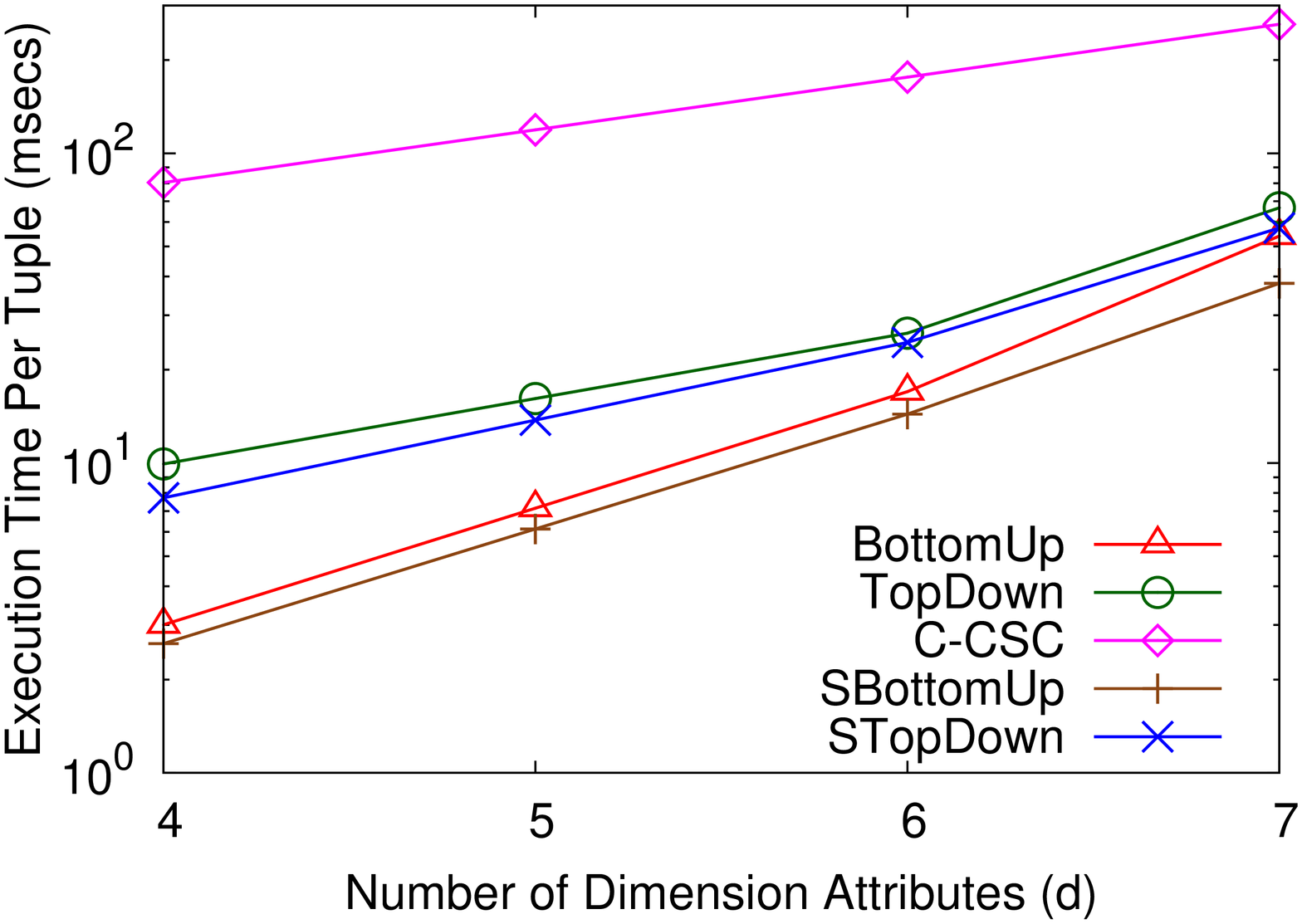, angle=-90, width=42mm,clip=}
   \caption{Varying $d$, $n$=$50{,}000$, $m$=$7$}
   \label{fig:100_m7}
\end{subfigure}
\hspace{1mm}
\begin{subfigure}[b]{.3\textwidth}
   \epsfig{file=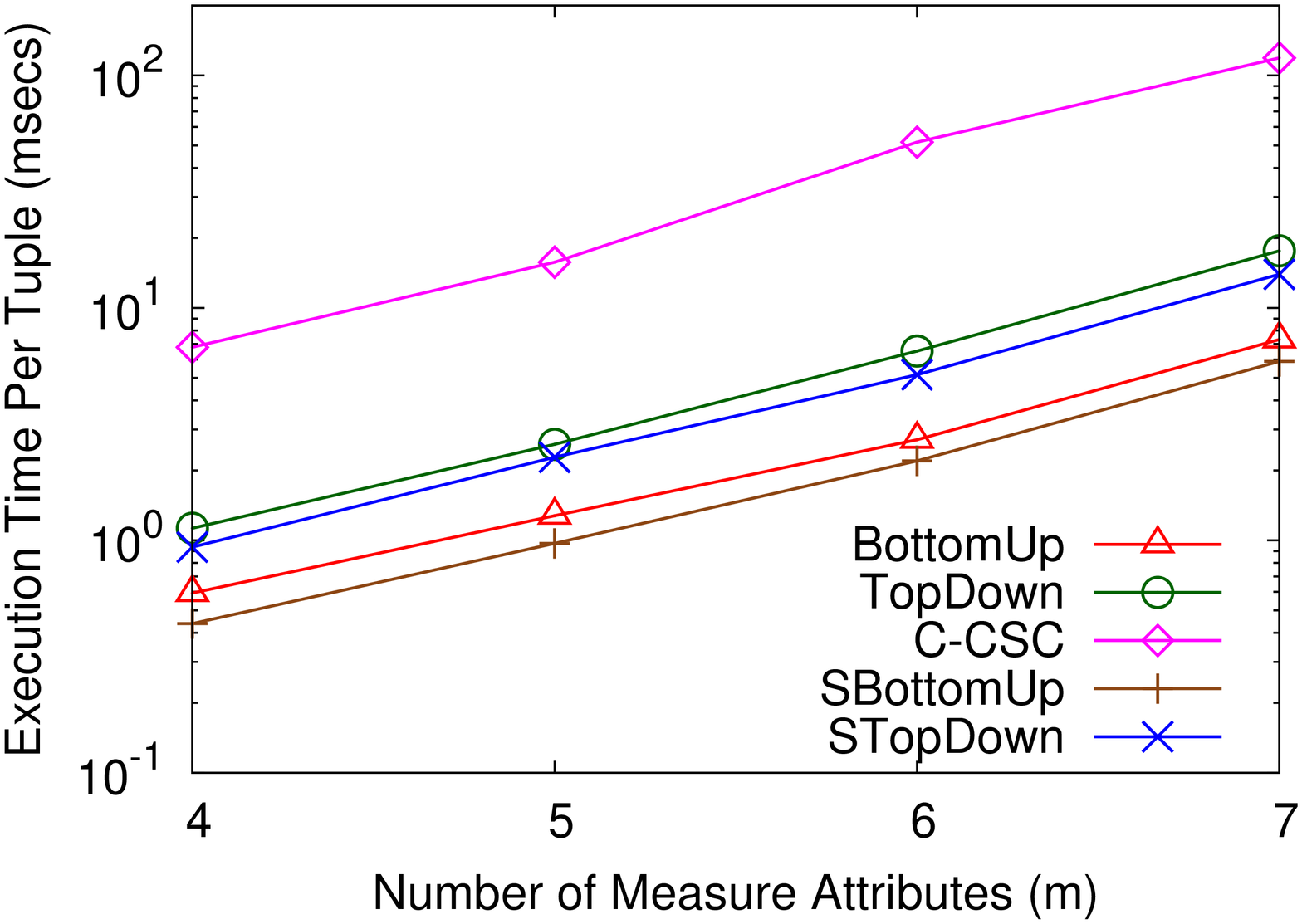, angle=-90, width=42mm,clip=}
   \caption{Varying $m$, $n$=$50{,}000$, $d$=$5$}
   \label{fig:100_d8}
\end{subfigure}\vspace{-2mm}
\hspace{1mm}
\caption{\small Execution Time of \algname{C-CSC}, \algname{BottomUp}, \algname{TopDown}, \algname{SBottomUp}, \algname{STopDown} on NBA dataset}
\label{fig:BUTDSBUSTD}
\end{minipage}
\hspace{-4mm}
\noindent \begin{minipage}{0.28\textwidth}
\centering
\epsfig{file=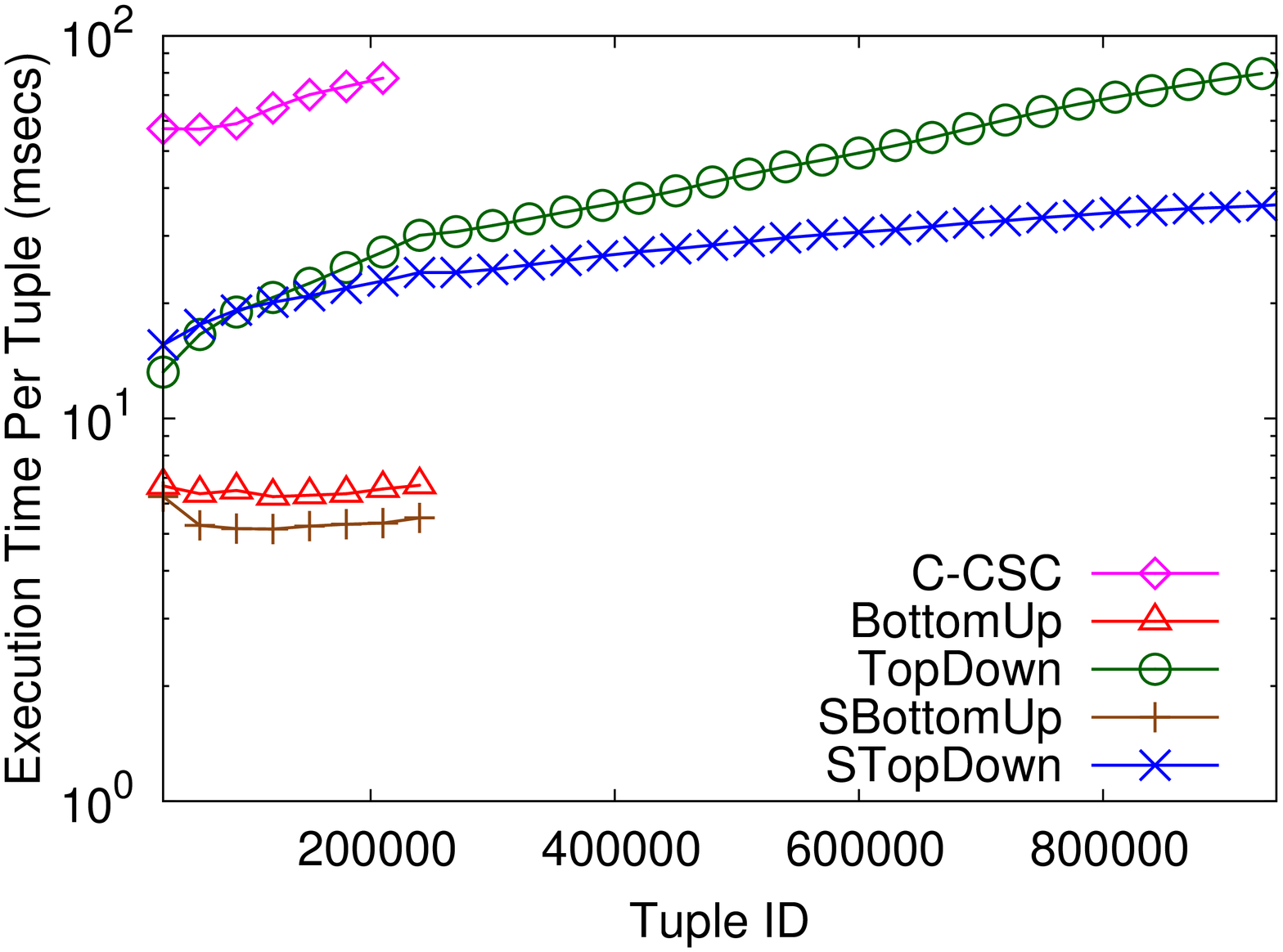, angle=-90, width=42mm,clip=}\vspace{-2mm}
\caption{\small Execution Time on the Weather Dataset, Varying $n$, $d$=$5$, $m$=$7$}
\label{fig:f14}
\end{minipage}
\end{figure*}

\textbf{Methods Compared}\hspace{2mm} We investigated
the performance of $7$ algorithms---the baseline algorithms \algname{BaselineSeq} and \algname{BaselineIdx}
from Sec.~\ref{sec:idea}, \algname{C-CSC} which is the \algname{CSC} adaptation described in Sec.~\ref{sec:related},
and the algorithms \algname{BottomUp}, \algname{TopDown}, \algname{SBottomUp} and
\algname{STopDown} from Sec.~\ref{sec:alg}.  We compared these
algorithms on both execution time and memory consumption.

\textbf{Parameters}\hspace{2mm} We ran our experiments
under combinations of five
parameters, which are number of dimension attributes ($d$), number of
measure attributes ($m$), number of tuples ($n$), maximum number of
bound dimension attributes ($\hat d$) and maximum number of measure
attributes allowed in measure subspaces ($\hat m$).
In Table~\ref{tab:d_list} (\ref{tab:t_list}), we list the dimension (measure) spaces
considered for different values of $d$ ($m$), which are subsets of the
aforementioned dimension (measure) attributes in the datasets.

\begin{table}[t]
\centering
\scriptsize
\begin{tabular}{|l|*{5}{c|}}\hline
$d$ & dimension space $\mathcal{D}$\\
\hline
\hline
$4$ & $\attrname{player},\ \attrname{season},\ \attrname{team},\ \attrname{opp\_team}$\\
\hline
$5$ & $\attrname{player},\ \attrname{season},\ \attrname{month},\ \attrname{team},\ \attrname{opp\_team}$\\
\hline
$6$ & $\attrname{position},\ \attrname{college},\ \attrname{state},\ \attrname{season},\ \attrname{team},\ \attrname{opp\_team}$\\
\hline
$7$ & $\attrname{position},\ \attrname{college},\ \attrname{state},\ \attrname{season},\ \attrname{month},\ \attrname{team},\ \attrname{opp\_team}$\\
\hline
\end{tabular}
\caption{\small Dimension Spaces for Different Values of $d$} 
\label{tab:d_list}
\end{table}

\begin{table}[t]
\centering
\scriptsize
\begin{tabular}{|l|*{5}{c|}}\hline
$m$ & measure space $\mathcal{M}$\\
\hline
\hline
$4$ & $\attrname{points},\ \attrname{rebounds},\ \attrname{assists},\ \attrname{blocks}$\\
\hline
$5$ & $\attrname{points},\ \attrname{rebounds},\ \attrname{assists},\ \attrname{blocks},\ \attrname{steals}$\\
\hline
$6$ & $\attrname{points},\ \attrname{rebounds},\ \attrname{assists},\ \attrname{blocks},\ \attrname{steals},\ \attrname{fouls}$\\
\hline
$7$ & $\attrname{points},\ \attrname{rebounds},\ \attrname{assists},\ \attrname{blocks},\ \attrname{steals},\ \attrname{fouls}$,\ \attrname{turnovers}\\
\hline
\end{tabular}
\caption{\small Measure Spaces for Different Values of $m$}
\label{tab:t_list}
\end{table}

In particular dimension/measure spaces (corresponding to $d$/$m$ values),
experiments were done for varying $\hat d$ and $\hat m$ values.
A constraint with more bound dimension attributes represents a more specific
context.  Similarly, a measure subspace with more measure attributes is more
specific.  Considering all possible constraint-measure pairs may thus produce
many over-specific and uninteresting facts.
The parameters $\hat d$ and $\hat m$ are for avoiding trivial facts.
For instance, if $d$=$5$, $m$=$4$, $\hat d$=$2$ and $\hat m$=$3$,
we consider all constraints with at most $2$ (out of $5$) bound dimension
attributes and all measure subspaces with at most $3$ (out of $4$) measure attributes.
In all experiments in this section, we set $\hat d = 4$ and
$\hat m = m$.  That means a constraint is allowed
to have up to $4$ bound attributes and a measure subspace can be any
subspace of the whole space $\mathcal{M}$ including $\mathcal{M}$
itself.  In Sec.~\ref{sec:interestingness}, we further study how
prominence of facts varies by $\hat d$ and $\hat m$ values. 

\subsection{Results of Memory-Based Implementation}\label{sec:memoryexp}
\begin{figure*}[t]
\noindent
\begin{minipage}{0.48\textwidth}
\centering
\begin{subfigure}[b]{.49\linewidth}
\centering

   \epsfig{file=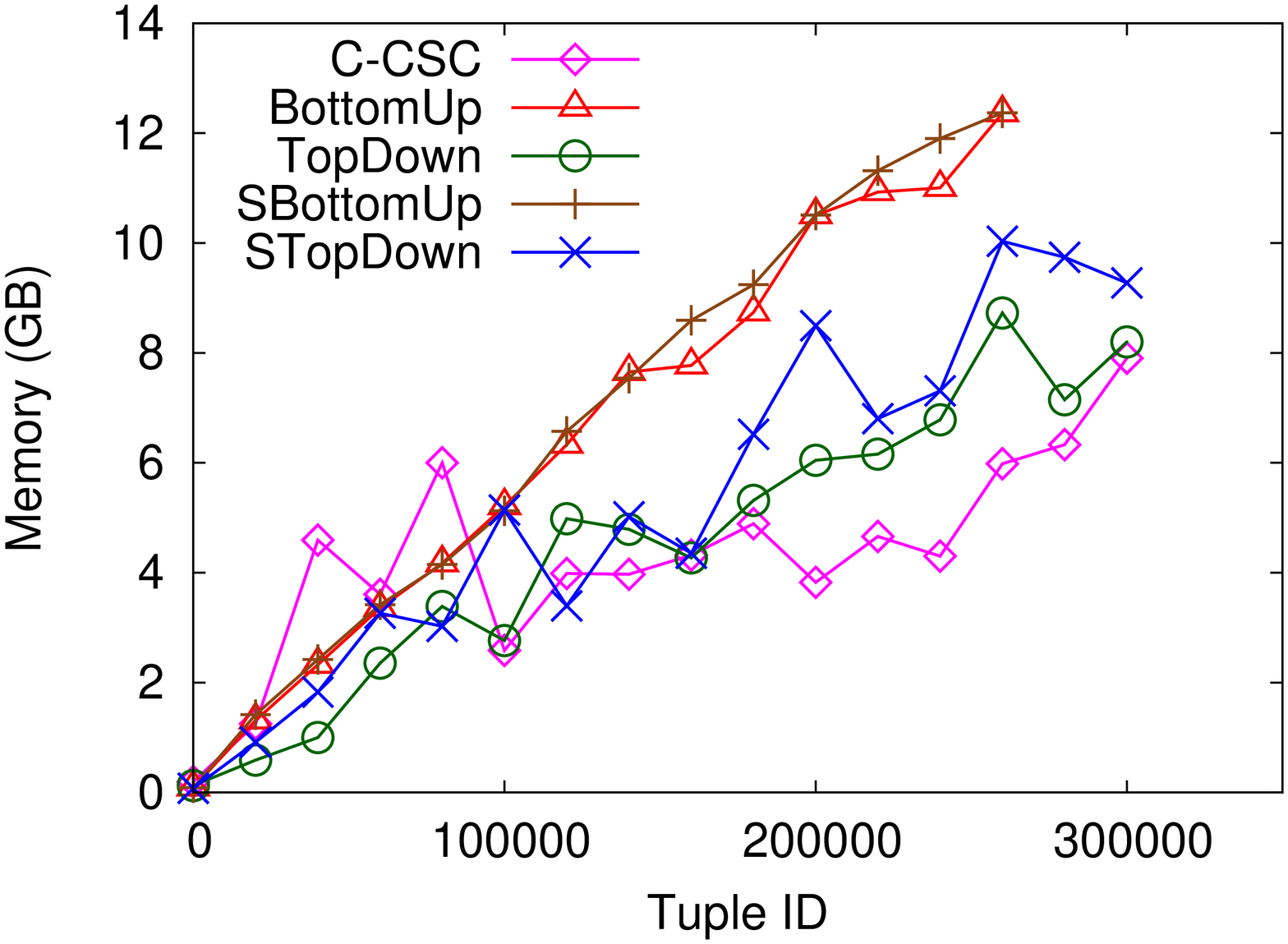, angle=-90, width=40mm,clip=}
   \caption{Size of Consumed Memory}
   \label{fig:8_7_-1_-1_memory}
\end{subfigure}
\begin{subfigure}[b]{.49\linewidth}
\centering
   \epsfig{file=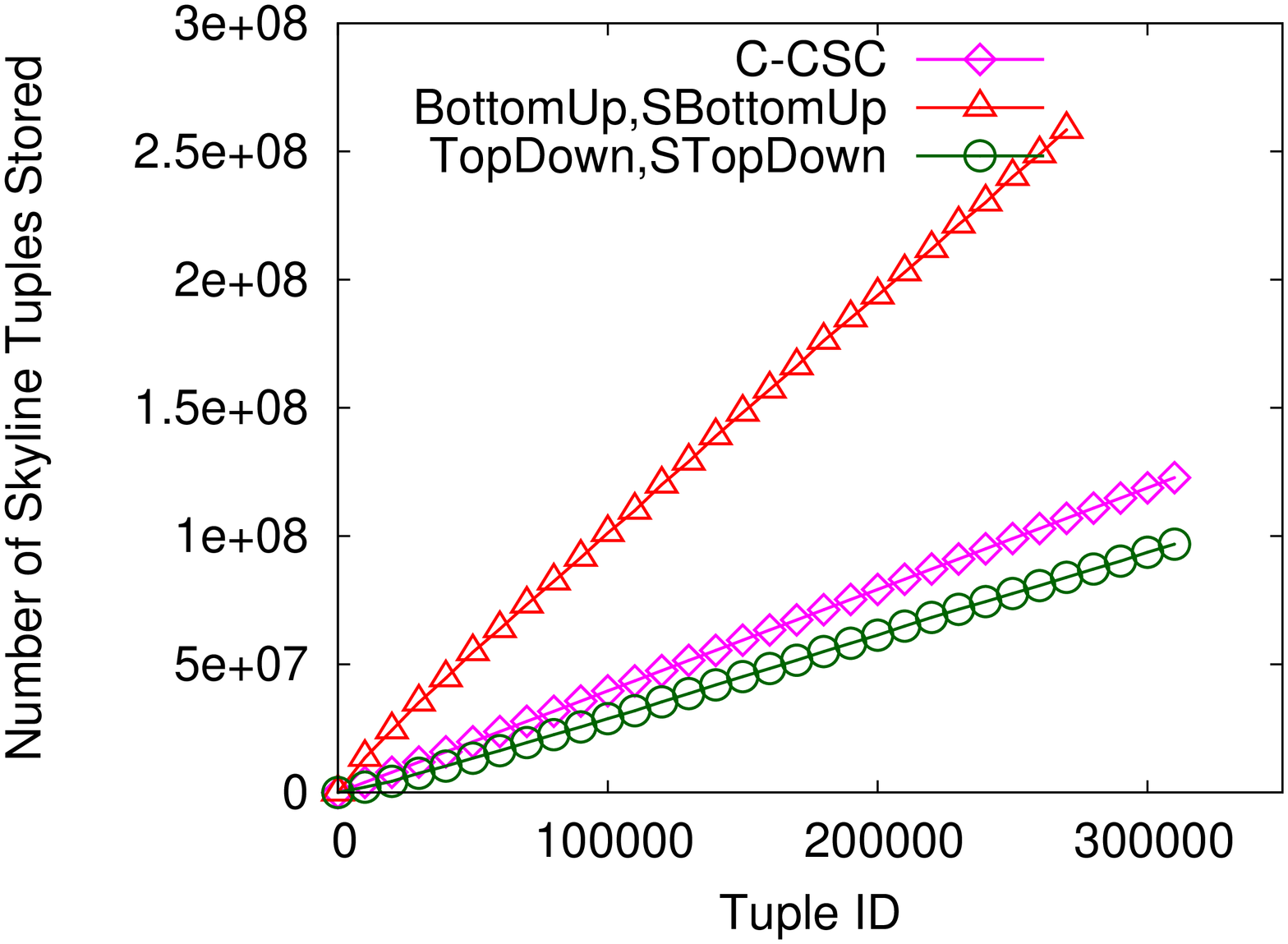, angle=-90, width=40mm,clip=}
   \caption{Num of Skyline Tuples Stored}
   \label{fig:8_7_300000_-1_-1}
\end{subfigure}
   \vspace{-6mm}
\caption{\small Memory Consumption by \algname{C-CSC}, \algname{BottomUp}, \algname{TopDown}, \algname{SBottomUp}, \algname{STopDown} on the NBA Dataset, Varying $n$, $d$=$5$, $m$=$7$}
\label{fig:memory}
\end{minipage}
\hspace{4mm}
\noindent
\begin{minipage}{0.48\textwidth}
\centering
\begin{subfigure}[b]{.49\linewidth}
\centering
   \epsfig{file=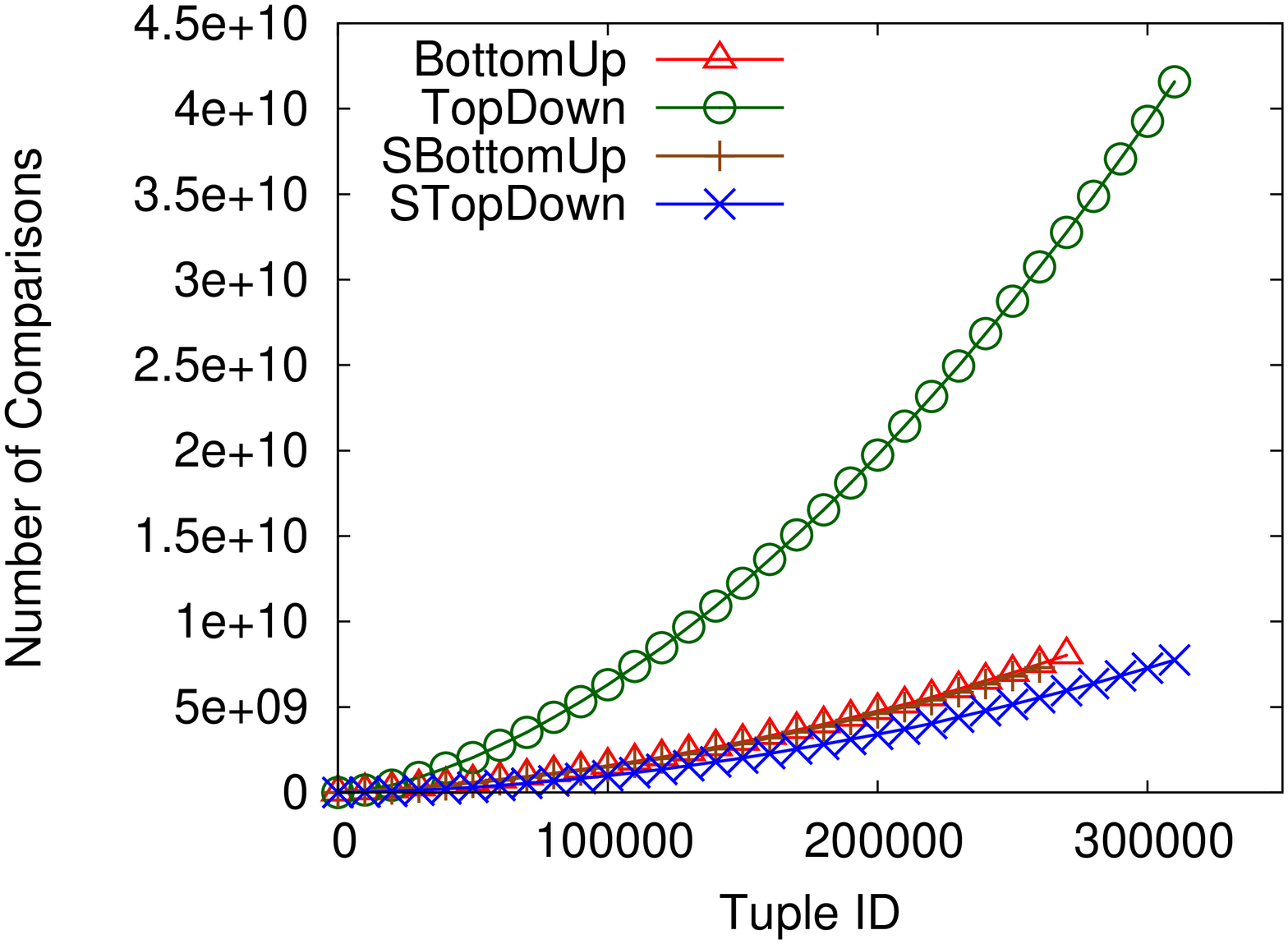, angle=-90, width=40mm,clip=}
   \caption{Number of Comparisons}
   \label{fig:c_id_5_7}
\end{subfigure}
\begin{subfigure}[b]{.49\linewidth}
\centering
   \epsfig{file=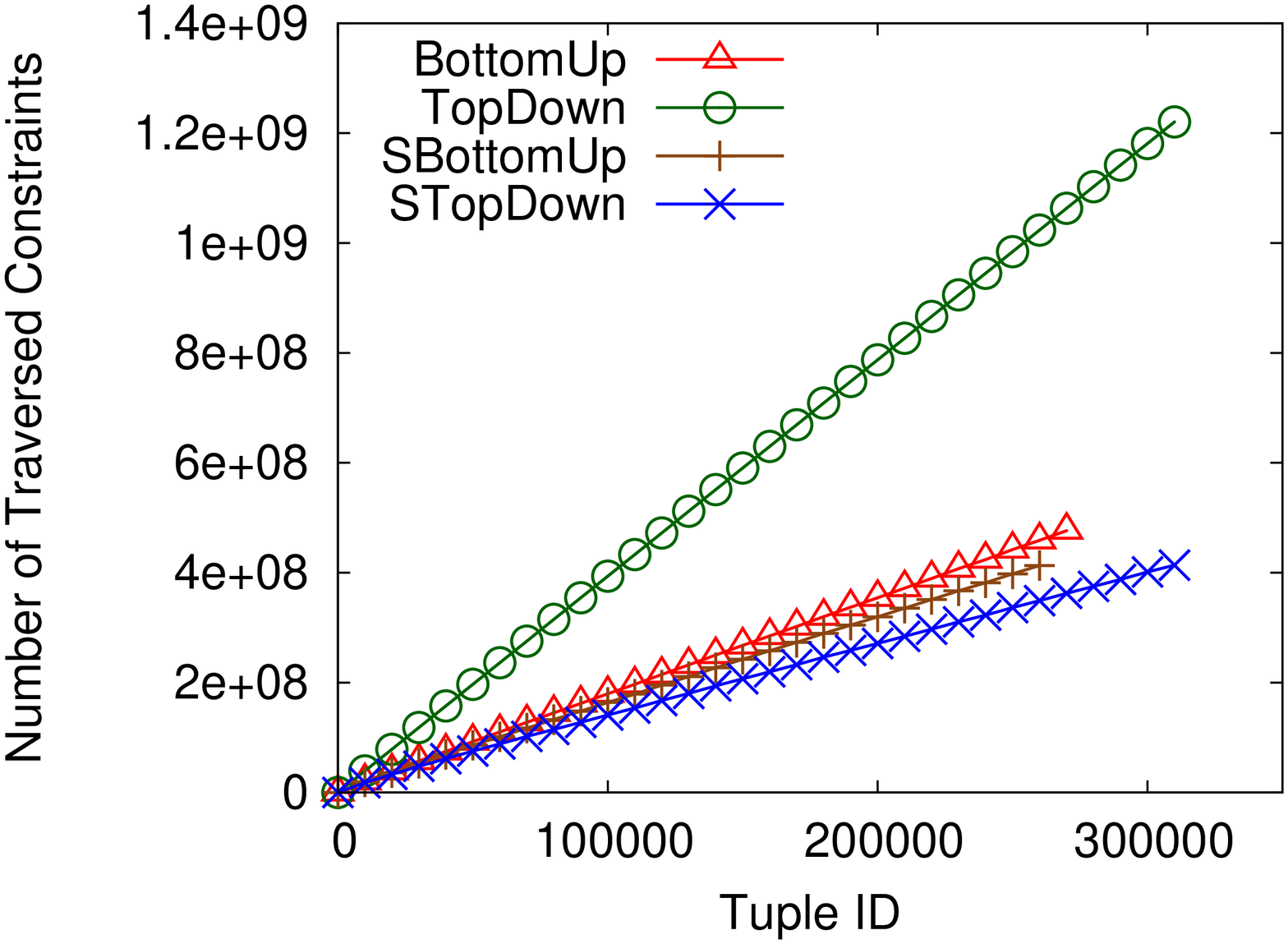, angle=-90, width=40mm,clip=}
   \caption{Num of Traversed Constraints}
   \label{fig:t_id_5_7}
\end{subfigure}
   \vspace{-6mm}
\caption{\small Work Done by \algname{BottomUp}, \algname{TopDown}, \algname{SBottomUp} and \algname{STopDown} on the NBA Dataset, Varying $n$, $d$=$5$, $m$=$7$}
\label{fig:work}
\end{minipage}
\end{figure*}

\begin{figure*}[t]
\hspace{-4mm}
\noindent \begin{minipage}{0.74\textwidth}
\centering
\begin{subfigure}[b]{.3\textwidth}
   \epsfig{file=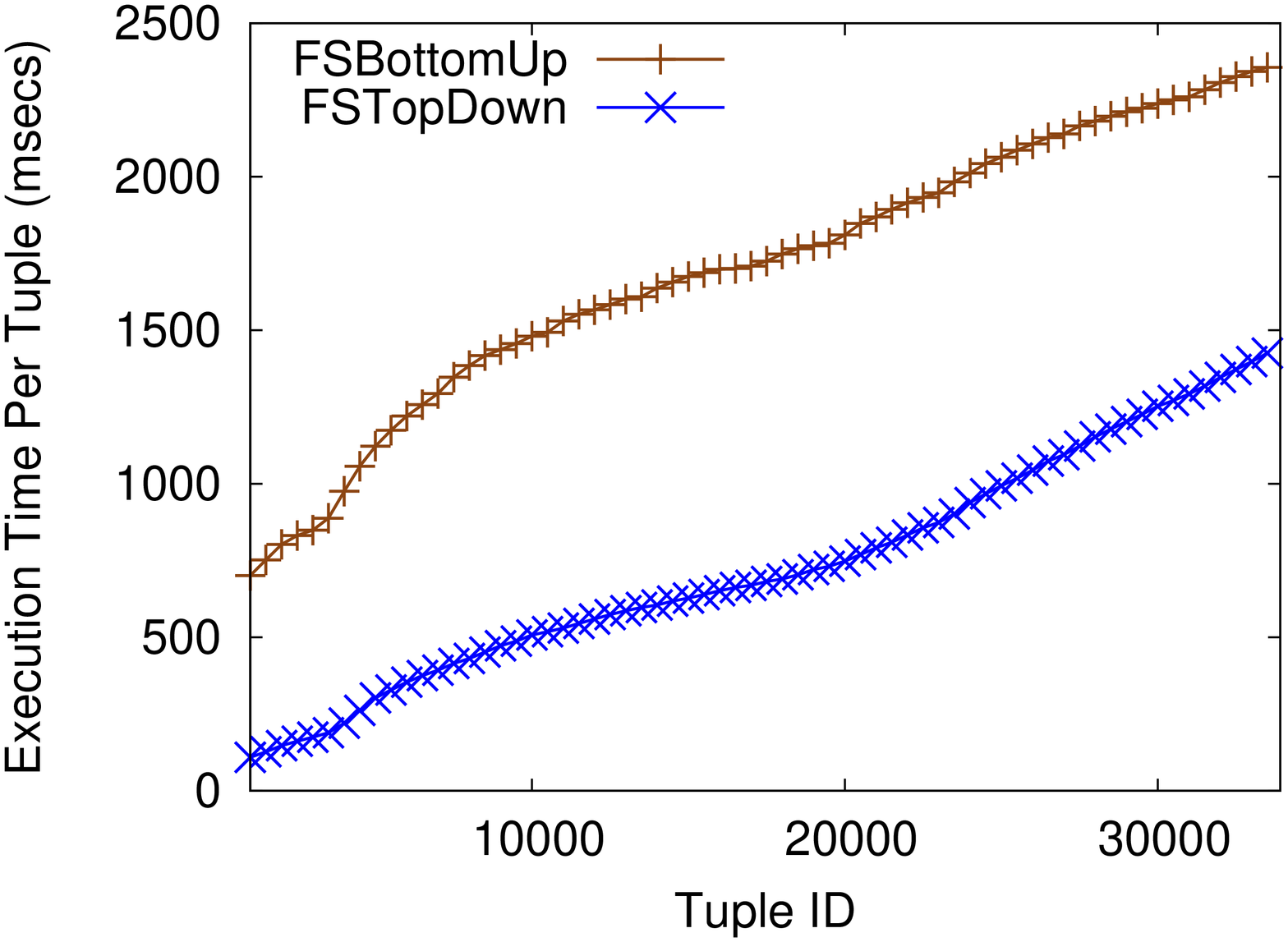, angle=-90, width=42mm,clip=}
   \caption{Varying $n$, $d$=$5$, $m$=$7$}
   \label{fig:time_file_8_7_-1_-1}
\end{subfigure}
\begin{subfigure}[b]{.3\textwidth}
   \epsfig{file=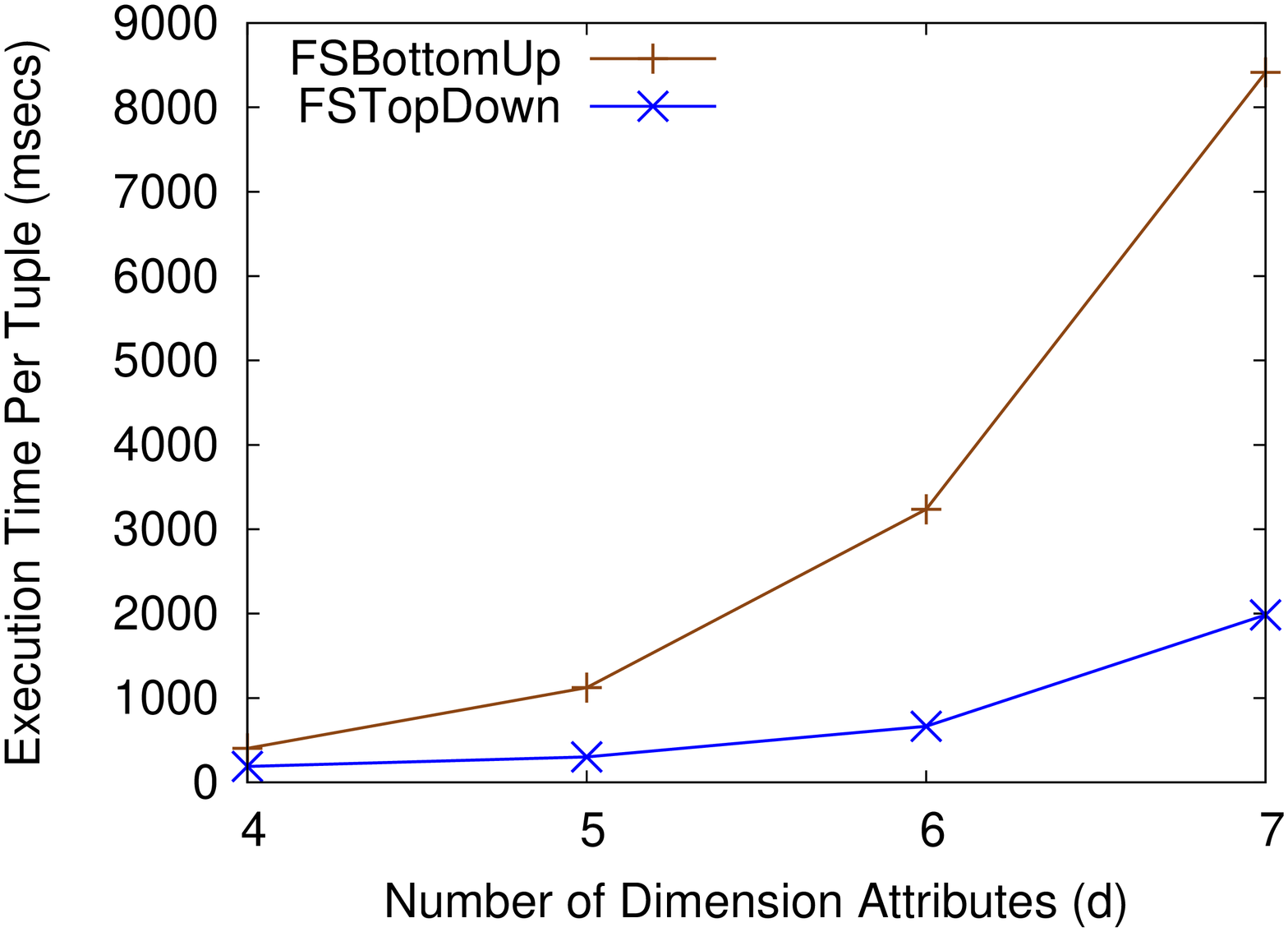, angle=-90, width=42mm,clip=}
   \caption{Varying $d$, $n$=$5{,}000$, $m$=$7$}
   \label{fig:file_1000_m5}
\end{subfigure}
\begin{subfigure}[b]{.3\textwidth}
   \epsfig{file=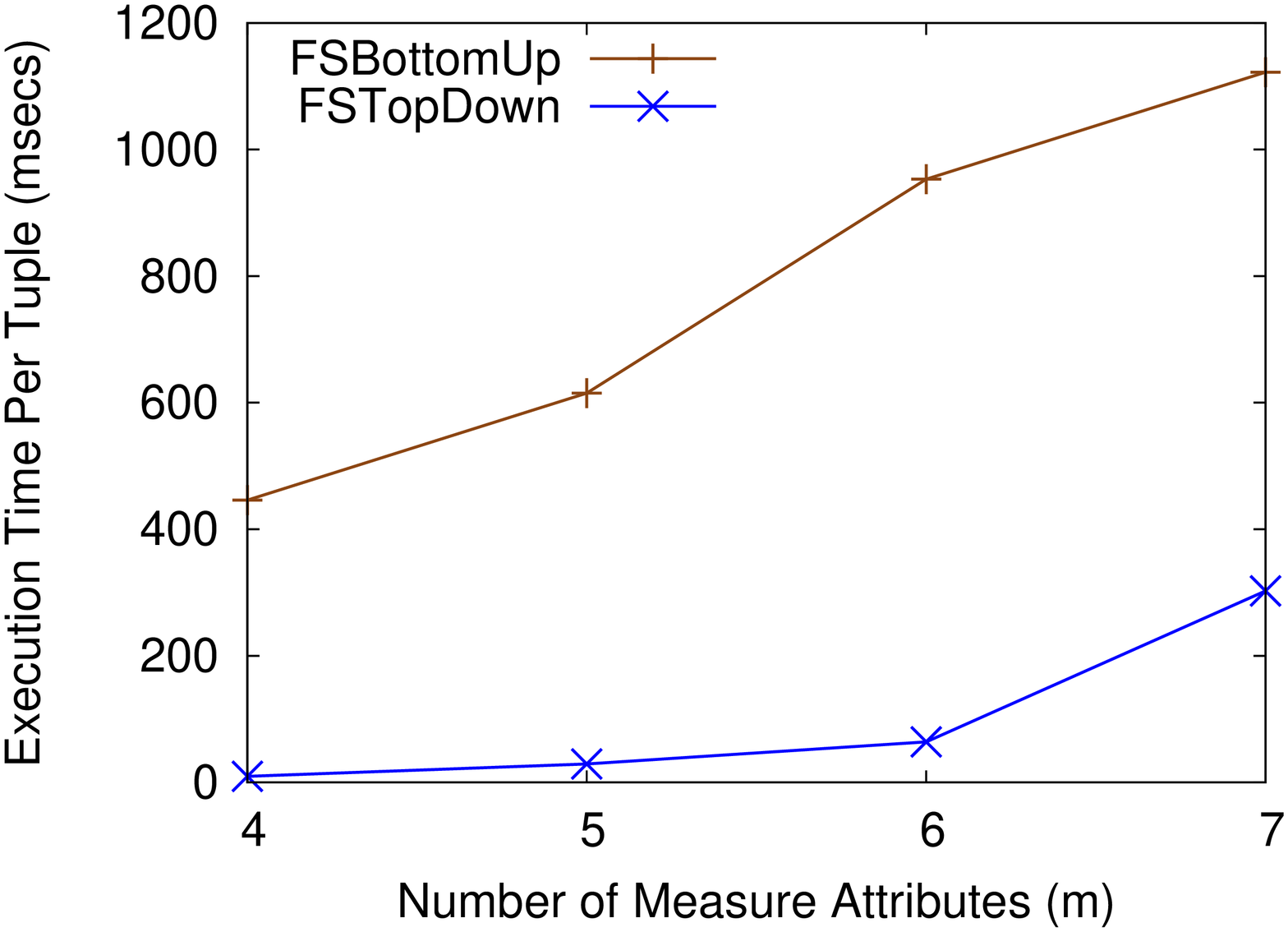, angle=-90, width=42mm,clip=}
   \caption{Varying $m$, $n$=$5{,}000$, $d$=$5$}
   \label{fig:file_1000_d5}
\end{subfigure}\vspace{-1mm}
\caption{\small Execution Time of \algname{FSBottomUp} and \algname{FSTopDown} on the NBA Dataset}
\label{fig:filebased}
\end{minipage}
\noindent \begin{minipage}{0.27\textwidth}
\centering
\epsfig{file=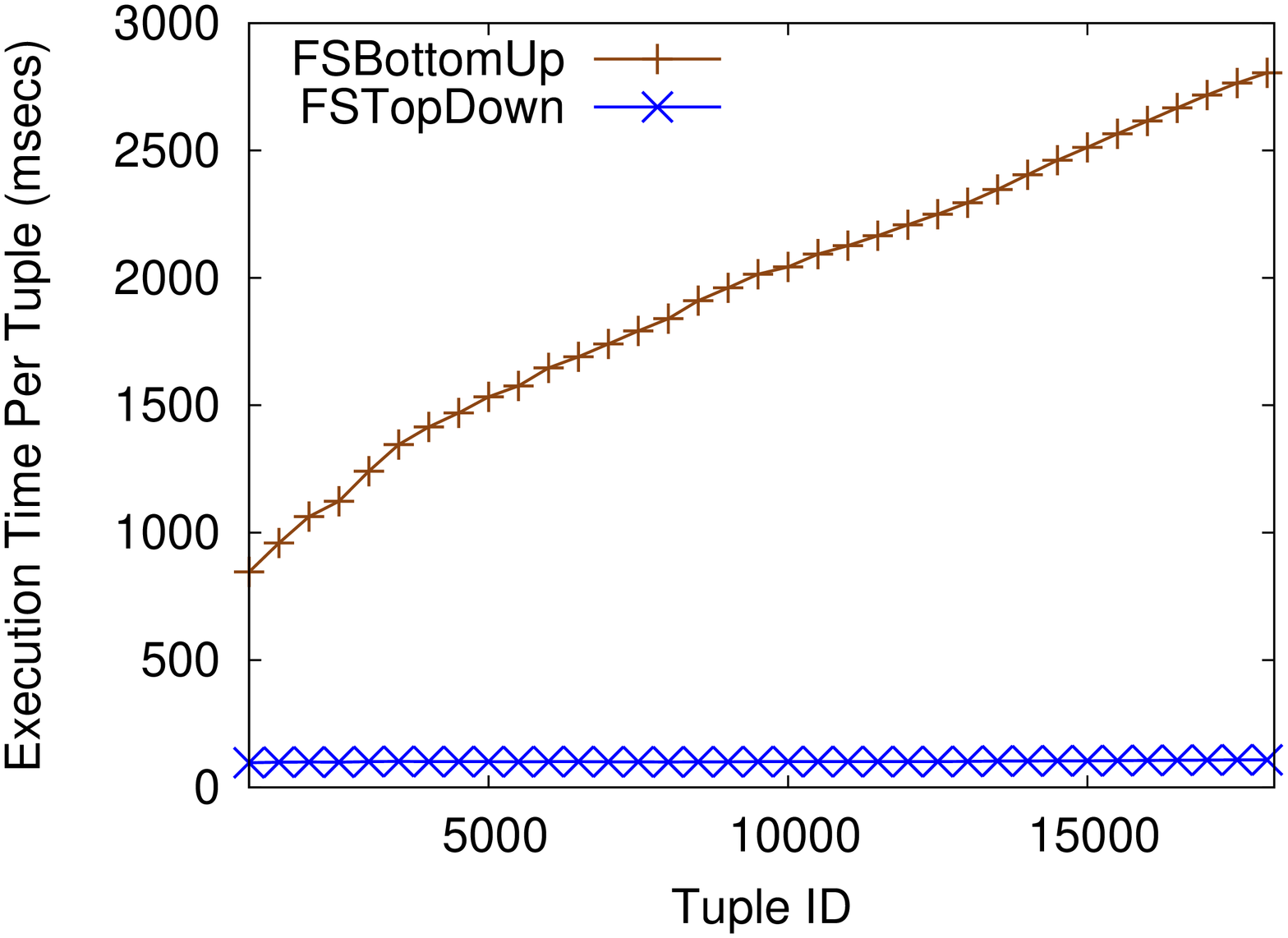, angle=-90, width=42mm,clip=}\vspace{-2mm}
\caption{\small Execution Time of \algname{FSBottomUp} and \algname{FSTopDown} on the Weather Dataset, Varying $n$, $d$=$5$, $m$=$7$}
\label{fig:weather_file}
\end{minipage}
\vspace{-3mm}
\end{figure*}

Fig.\ref{fig:BaselineBUTD} compares the per-tuple execution times (by milliseconds, in logarithmic scale) of \algname{BaselineSeq}, \algname{BaselineIdx}, \algname{C-CSC}, \algname{BottomUp} and \algname{TopDown} on the NBA dataset.  Fig.\ref{fig:3_3_-1_-1_MI_BI_BU_TD} shows how the per-tuple execution times increase as the algorithms process tuples sequentially by their timestamps.  The values of $d$ and $m$ are $d$=$5$ and $m$=$7$.  Fig.\ref{fig:1000_m5} shows the times under varying $d$, given $n$=$50{,}000$ and $m$=$7$. Fig.\ref{fig:1000_d5} is for varying $m$, $n$=$50{,}000$ and $d$=$5$.   The figures demonstrate that \algname{BottomUp} and \algname{TopDown} outperformed the baselines by orders of magnitude and \algname{C-CSC} by one order of magnitude.  
Furthermore, Fig.\ref{fig:1000_m5} and Fig.\ref{fig:1000_d5} show that the execution time of all these algorithms increased exponentially by both $d$ and $m$, which is not surprising since the space of possible constraint-measure pairs grows exponentially by dimensionality.

Fig.\ref{fig:BUTDSBUSTD} uses the same configurations in
Fig.\ref{fig:BaselineBUTD} to compare \algname{C-CSC}, \algname{BottomUp},
\algname{TopDown}, \algname{SBottomUp} and \algname{STopDown}.  We
make the following observations on the results.  First, \algname{C-CSC}
was outperformed by one order of magnitude.  The per-tuple execution
times of all algorithms exhibited moderate growth with respect to
$n$ and superlinear growth with respect to $d$ and $m$, matching the
observations from Fig.\ref{fig:BaselineBUTD}.\vspace{-0.5mm}

Second, in Fig.\ref{fig:8_7_300000_3_-1}, the bottom-up algorithms
exhausted available JVM heap and were terminated due to memory
overflow before all tuples were consumed.  On the contrary, the
top-down algorithms finished all tuples.  This difference was more
clear on the larger weather dataset (Fig.\ref{fig:f14}), on which
the bottom-up algorithms caused memory overflow shortly after $0.2$
million tuples were encountered, while the top-down algorithms were
still running normally after $0.9$ million tuples.
As the difference was already clear after $0.9$ million tuples, we terminated
the executions of top-down algorithms at that point. 
The difference in the sizes of consumed memory by these two categories
of algorithms is shown in Fig.\ref{fig:8_7_-1_-1_memory}.
The difference in memory
consumption is due to that \algname{TopDown}/\algname{STopDown} only
store a skyline tuple at its maximal skyline constraints, while
\algname{BottomUp}/\algname{SBottomUp} store it at all skyline
constraints.  This observation is verified by
Fig.\ref{fig:8_7_300000_-1_-1}, which shows how the number of
stored skyline tuples increases by $n$.  We see that
\algname{BottomUp}/\algname{SBottomUp} stored several times more
tuples than \algname{TopDown}/\algname{STopDown}.  Note that
\algname{TopDown} and \algname{STopDown} use the same skyline
tuple materialization scheme.  Correspondingly \algname{BottomUp} and
\algname{SBottomUp} store tuples in the same way.\vspace{-0.5mm}

Fig.\ref{fig:f14} also shows that, for the weather dataset, \algname{C-CSC}
could not proceed shorty after 0.2 million tuples were processed.  This was
also due to memory overflow caused by \algname{C-CSC}, since it needs to store
skyline tuples in their ``minimum subspaces''.  \algname{C-CSC} did not exhaust memory
when it processed the NBA dataset (Fig.\ref{fig:8_7_300000_3_-1}), since there were
less skyline tuples in the smaller dataset.

Third, in terms of execution time,
\algname{TopDown}/\algname{STopDown} were outperformed by
\algname{BottomUp}/\algname{SBottomUp}.  The reason is, if a new
tuple $t$ dominates a previous tuple $t'$ in constraint $C$ and
measure subspace $M$, \algname{TopDown}/\algname{STopDown} must update
$\mathcal{MSC}^{t'}_M$. On the contrary,
\algname{BottomUp}/\algname{SBottomUp} do not carry this overhead;
they only need to delete $t'$ from $\mu_{C,M}$.  Thus, there is a
space-time tradeoff between the top-down and bottom-up
strategies.

Finally, \algname{SBottomUp}/\algname{STopDown} are faster than
\algname{BottomUp} / \algname{TopDown}, which is the benefit of sharing
computation across measure subspaces.  Figs.\ref{fig:100_m7}
and~\ref{fig:100_d8} show that this benefit became more prominent with the increase of both $d$ and
$m$.  Fig.\ref{fig:work} further presents the amount of work done
by these algorithms, in terms of compared tuples (Fig.\ref{fig:c_id_5_7})
and traversed constraints (Fig.\ref{fig:t_id_5_7}).
There are substantial differences between
\algname{TopDown} and \algname{STopDown}, but the differences between
\algname{BottomUp} and \algname{SBottomUp} are insignificant.  The
reason is as follows.  \algname{STopDown} avoids visiting pruned
non-skyline constraints, which \algname{TopDown} cannot avoid.
Although \algname{SBottomUp} avoids such non-skyline constraints too,
\algname{BottomUp} also avoids most of them.  The difference between
\algname{BottomUp} and \algname{SBottomUp} is that \algname{BottomUp}
still visits the boundary non-skyline constraints that are parents of
skyline constraints and then skips their ancestors, while
\algname{SBottomUp} skips all non-skyline constraints.  Such a
difference on boundary non-skyline constraints is not significant.

\vspace{-1mm}
\subsection{Results of File-Based Implementation}\label{sec:fileexp}
\vspace{-1mm}
The memory-based implementations of all algorithms store skyline tuples for all combinations of constraints and measure subspaces.
As a dataset grows, sooner or later, all algorithms will lead to memory overflow. 
To address this, we investigated file-based implementations of \algname{STopDown} and \algname{SBottomUp}, denoted \algname{FSTopDown} and \algname{FSBottomUp}, respectively.
We did not include \algname{C-CSC} in this experiment since Figs.\ref{fig:BaselineBUTD}-\ref{fig:memory} clearly
show \algname{TopDown}/\algname{STopDown} is one order of magnitude faster than \algname{C-CSC} and consumes about the same amount of memory.

In the file-based implementations, each non-empty $\mu_{C,M}$ is stored as a binary file.  Since the size of $\mu_{C,M}$ for any particular constraint-measure pair $(C,M)$ is small, all tuples in the corresponding file are read into a memory buffer when the pair is visited.  Insertion and deletion on $\mu_{C,M}$ are then performed on the buffer.  When an algorithm finishes process the pair, the file is overwritten by the buffer's content.

Fig.\ref{fig:filebased} uses the same configurations in Figs.\ref{fig:BaselineBUTD} and~\ref{fig:BUTDSBUSTD} to compare the per-tuple execution times of \algname{FSBottomUp} and \algname{FSTopDown} on the NBA dataset.  Fig.\ref{fig:weather_file} further compares them on the weather dataset.   The figures show that \algname{FSTopDown} outperformed \algname{FSBottomUp} by multiple times.  Even for only $n$=$5{,}000$, their performance gap was already clear in Figs.\ref{fig:file_1000_m5} and~\ref{fig:file_1000_d5}.  The reason is as follows.  In file-based implementation, while traversing a pair $(C,M)$, a file-read operation occurs if $\mu_{C,M}$ is non-empty.  Since \algname{FSTopDown} stores significantly fewer tuples than \algname{FSBottomUp} (cf. Fig.\ref{fig:memory}), \algname{FSTopDown} is more likely to encounter empty $\mu_{C,M}$ and thus triggers fewer file-read operations.  Further, a file-write operation occurs if the algorithms must update $\mu_{C,M}$.  Again, since \algname{FSTopDown} stores fewer tuples, it requires fewer file-write operations.  Hence, although \algname{SBottomUp} outperformed \algname{STopDown} on in-memory execution time, \algname{FSTopDown} triumphed \algname{FSBottomUp} because I/O-cost dominates in-memory computation.

\section{Case Study}\label{sec:interestingness}

\begin{figure*}[t]
\noindent \begin{minipage}{0.26\textwidth}
\centering
    \epsfig{file=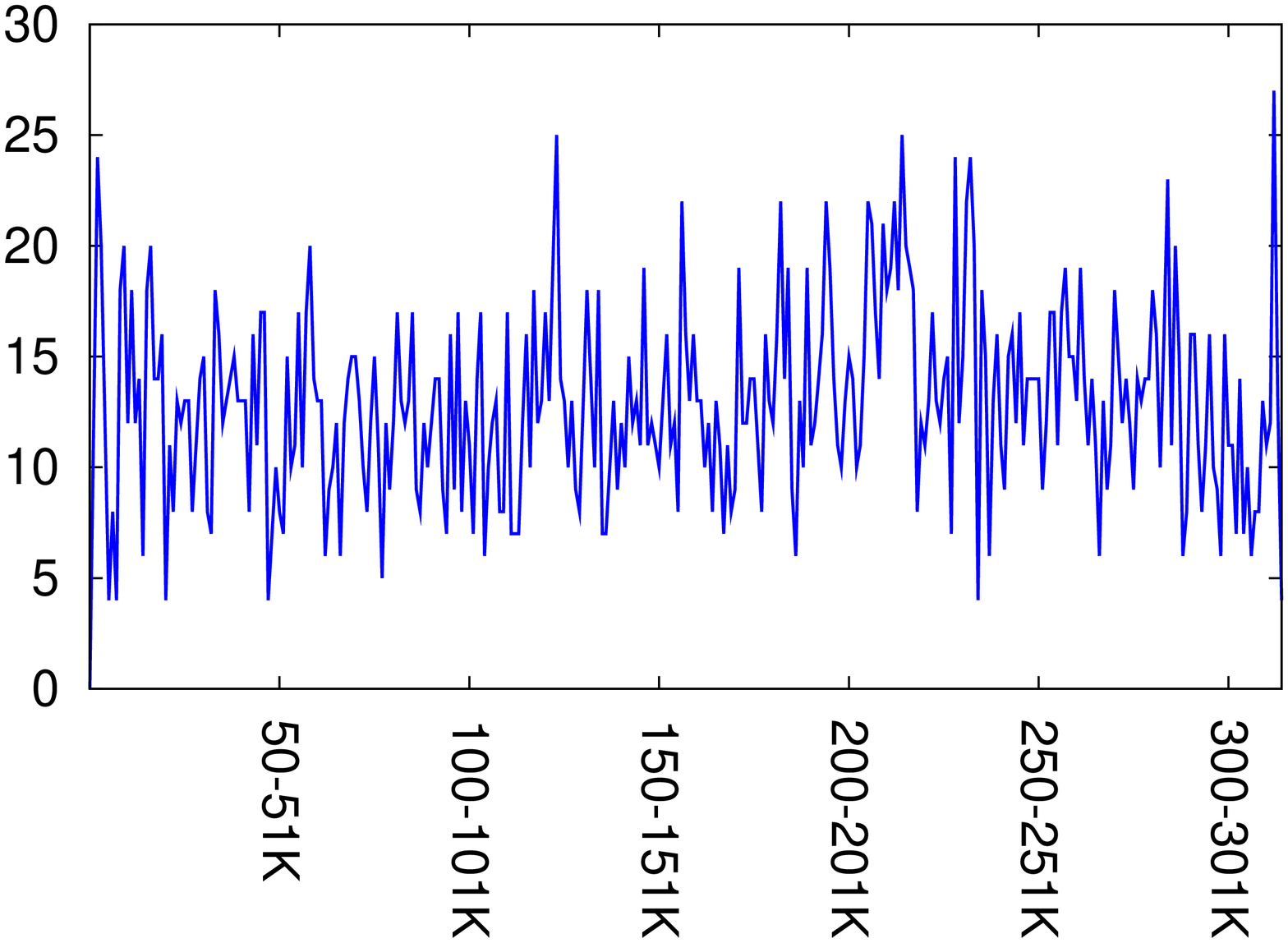, angle=-90, width=45mm,clip=}\vspace{-2mm}
    \caption{\small Number of Prominent Facts for Each $1$K Tuples, $\tau$=$10^3$}
    \label{fig:5_7_1_3_1000_0.001}
\end{minipage}
\hspace{6mm}
\noindent \begin{minipage}{.74\textwidth}
\begin{subfigure}[b]{.47\textwidth}
\centering
    \vspace{-3mm}\epsfig{file=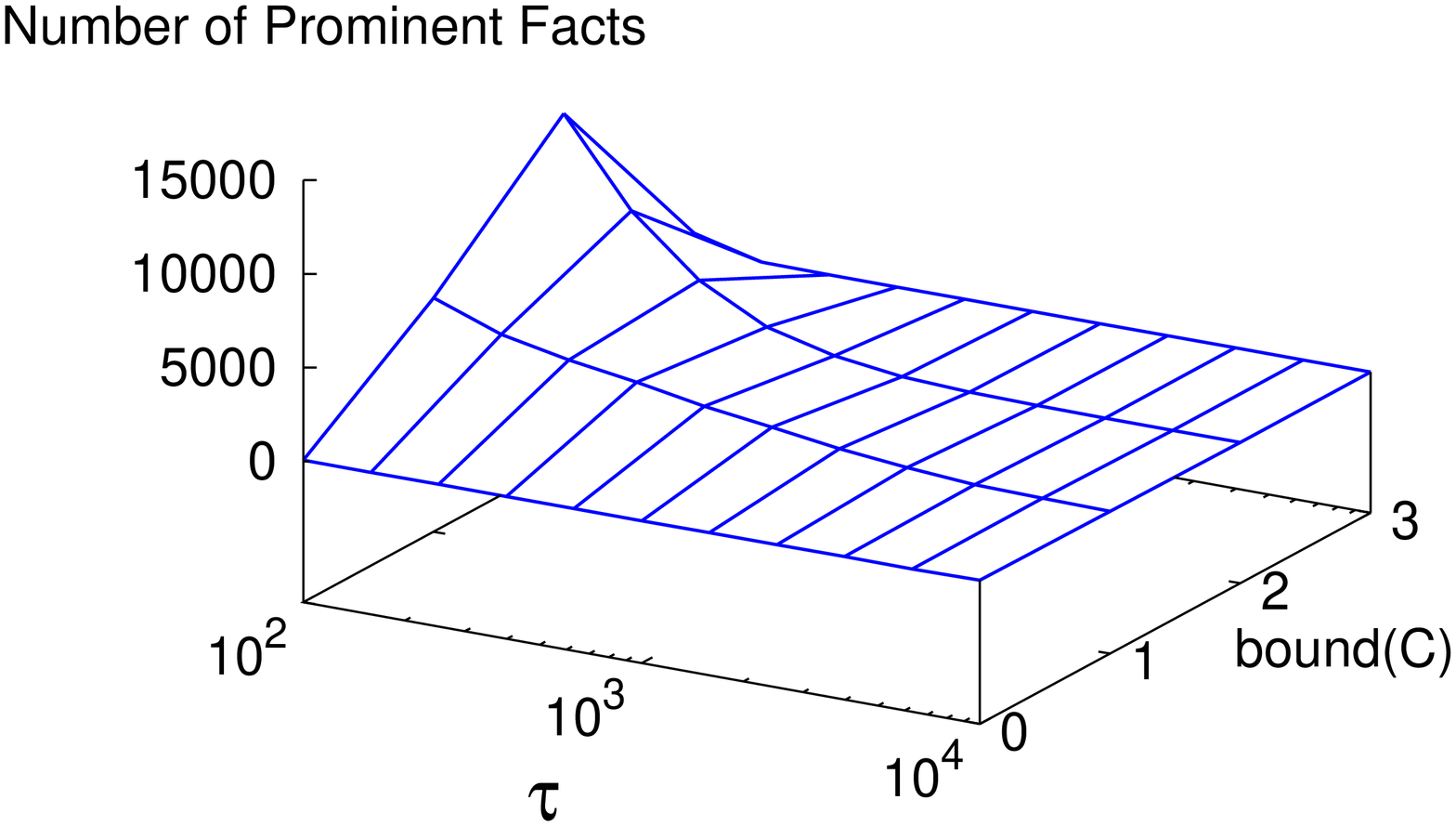, angle=-90, width=57mm,clip=}\vspace{-3mm}
    \caption{By Number of Bound Dimension Attributes}
    \label{fig:5_7_1_3_D.001}
\end{subfigure}
\begin{subfigure}[b]{.47\textwidth}
\centering
    \vspace{-3mm}\epsfig{file=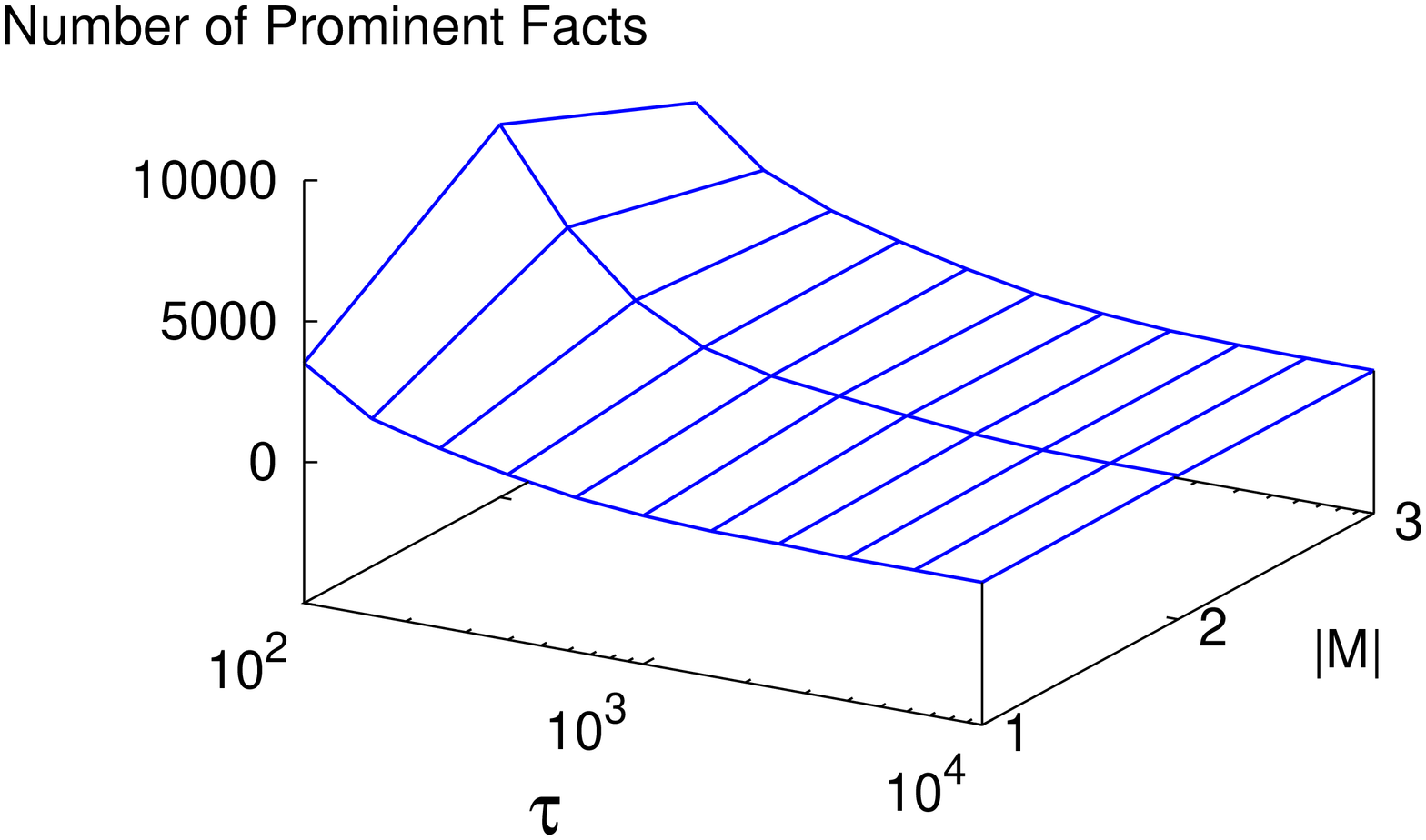, angle=-90, width=57mm,clip=}\vspace{-3mm}
    \caption{By Dimensionality of Measure Subspaces}
    \label{fig:5_7_1_3_M.001}
\end{subfigure}
\vspace{-2mm}\caption{\small Distribution of Prominent Facts, Varying $\tau$}
\label{fig:num_fact}
\end{minipage}
\vspace{-3mm}
\end{figure*}

A tuple may be in the contextual skylines of many
constraint-measure pairs.  For instance, $t_7$ in
Example~\ref{ex:motivate} belongs to $196$ contextual skylines (of
course partly because the table is tiny and most contexts
contain only $t_7$).  Reporting all such facts
overwhelms users and makes important facts harder to spot.  It is
crucial to report truly \emph{prominent} facts, which should be rare.
We measure the \emph{prominence}
of a fact (i.e., a constraint-measure pair $(C,M)$) by $\smash{|\sigma_C(R)| \over |\lambda_M(\sigma_C(R))|}$, the
cardinality ratio of all tuples to skyline tuples in the context.
Consider two pairs in Example~\ref{ex:motivate}:($C_1$:\attrname{month}=\attrval{Feb},$M_1$:\{\attrnametwo{points},\attrnametwo{assists},\attrnametwo{rebounds}\}) and
($C_2$:\attrname{team}=\attrval{Celtics}$\wedge$\attrname{opp\_team}=\attrval{Nets},$M_2$:\{\attrnametwo{assists},\attrnametwo{rebounds}\}).
The context of $C_1$ contains $5$ tuples,
among which $t_2$ and $t_7$ are in the skyline in $M_1$.
Hence, the prominence of ($C_1,M_1$) is $5/2$.  Similarly the
prominence of ($C_2,M_2$) is $3/2$. Hence ($C_1,M_1$) is more
prominent, because larger ratios indicate rarer events.

For a newly arrived tuple $t$, we rank all situational facts $S^t$ pertinent to $t$ in descending order of their prominence.  A fact is \emph{prominent} if its prominence value is the highest among $S^t$ and is not below a given threshold $\tau$.  (There can be multiple prominent facts pertinent to the arrival of $t$, due to ties in their prominence values.)   Consider $t_7$ in Example~\ref{ex:motivate}.  From the $196$ facts in $S^{t_7}$, the highest prominence value is $3$.  If $\tau$$\leq$$3$, those facts in $S^{t_7}$ attaining value $3$ are the prominent facts pertinent to $t_7$.  Among many such facts, examples are (\attrname{player}=\attrval{Wesley}, \{\attrnametwo{rebounds}\}) and (\attrname{month}=\attrval{Feb.}$\wedge$\attrname{team}=\attrval{Celtics},\{\attrnametwo{points}\}).
Note that, based on the definition of the prominence measure and the threshold $\tau$, a context must have at least $\tau$ tuples in order to contribute a prominent fact.



We studied the prominence of situational facts from the NBA dataset,
under the parameter setting $d$=$5$, $m$=$7$, $\hat d$=$3$, $\hat
m$=$3$ and $\tau$=$500$.  In other words, each prominent fact on a new
tuple $t$ is about a contextual skyline that contains $t$ and at most
$0.2\%$ of the tuples in the context.  Below we show some of the
discovered prominent facts.  They do not necessarily
stand in the real world, since our dataset does not include the
complete NBA records from all seasons. 
\begin{list}{$\bullet$}
{ \setlength{\leftmargin}{1em} }
\item Lamar Odom had 30 points, 19 rebounds and 11 assists on March 6, 2004. No one before had a better or equal performance in NBA history.
\item Allen Iverson had 38 points and 16 assists on April 14, 2004 to become the first player with a 38/16 (points/assists) game in the 2004-2005 season.
\item Damon Stoudamire scored 54 points on January 14, 2005. It is the highest score in history made by any Trail Blazers.
\end{list}

Figs.\ref{fig:5_7_1_3_1000_0.001}
and~\ref{fig:num_fact} help us further understand the prominent facts
from this experiment at the macro-level.
Fig.\ref{fig:5_7_1_3_1000_0.001} shows the number of prominent
facts for each 1000 tuples, given threshold $\tau$=$10^3$.  For
instance, there are $11$ prominent facts in total from the
$100{,}000^{\text{th}}$ tuple to the $101{,}000^{\text{th}}$ tuple.
We observed that the values in Fig.\ref{fig:5_7_1_3_1000_0.001}
mostly oscillate between $5$ and $25$.  Consider the number of tuples
and the huge number of constraint-measure pairs, these
prominent facts are truly selective.  One might expect a downward
trend in Fig.\ref{fig:5_7_1_3_1000_0.001}. It did
not occur due to the constant formulation of new contexts.  Each year, a
new NBA regular season commences and some new players start to play.
Such new values of dimension attributes
\attrname{season} and \attrname{player}, coupled with combinations of
other dimension attributes, form new contexts.  Once a context is
populated with enough tuples (at least $\tau$), a newly arrived tuple
belonging to the context may trigger a prominent fact.

Fig.\ref{fig:5_7_1_3_D.001} shows the distribution of prominent
facts by the number of bound dimension attributes in constraint for
varying $\tau$ in $[10^2, 10^4]$.
Fig.\ref{fig:5_7_1_3_M.001} shows the distribution by the
dimensionality of measure subspace.  We observed fewer
prominent facts with $0$ and $3$ bound attributes (out of $d$=$5$
dimension attributes) than those with $1$ and $2$ bound attributes,
and fewer prominent facts in measure subspaces with $1$ and
$3$ attributes than those with $2$ attributes.
The reasons are: \textbf{1)}~With regard to dimension
attributes, if there are no bound attributes in the constraint, the
context includes the whole table.  Naturally it is more challenging to
establish a prominent fact for the whole table.  If the constraint has
more bound attributes, the corresponding context becomes more specific
and contains fewer tuples, which may not be enough to contribute a
prominent fact (recall that having one prominent fact entails
a context size of no less than $\tau$).  Therefore, there are fewer
prominent streaks with $3$ bound attributes.  \textbf{2)}~With regard
to measure attributes, on a single measure, a tuple must have the
highest value in order to top other tuples, which does not often
happen.  There are thus fewer prominent facts in single-attribute
subspaces.  In a subspace with $3$ attributes, there are also fewer
prominent facts, because the contextual skyline contains more tuples,
leading to a smaller prominence value that may not beat the threshold
$\tau$.


\section{Conclusion}\label{sec:conclusion}
We studied the novel problem of discovering prominent situational facts, which is formalized as finding the constraint-measure pairs that qualify a new tuple as a contextual skyline tuple.  We presented algorithms for efficient discovery of prominent facts.  We used a simple prominence measure to rank discovered facts. Extensive experiments over two real datasets validated the effectiveness and efficiency of the techniques. This is our first step towards general fact finding for computational journalism. Going forward, we plan to explore several  directions, including generalizing the solution for allowing deletion and update of data, narrating facts in natural-language text and reporting facts of other forms (e.g., facts about multiple tuples in a dataset and aggregates over tuples).

\section{Acknowledgement}\label{sec:acknowledgement}

The work of Li is partially supported by NSF Grant IIS-1018865, CCF-1117369, 2011 and 2012 HP Labs Innovation Research Award, and the National Natural Science Foundation of China Grant 61370019.  The work of Yang is supported by IIS-0916027 and IIS-1320357. Any opinions, findings, and conclusions or recommendations expressed in this publication are those of the author(s) and do not necessarily reflect the views of the funding agencies.

\bibliographystyle{abbrv}
\small
\bibliography{factmonitoring}
\normalsize

\end{document}